\magnification=1200

\hsize 17truecm \vsize 23truecm

\font\twelvec=msbm10 at 12pt \font\sevenc=msbm10 at 9pt
\font\fivec=msbm10 at 7pt

\newfam\co
\textfont\co=\twelvec \scriptfont\co=\sevenc
\scriptscriptfont\co=\fivec

\def\arg{\mathop{\rm arg}\nolimits}
\def\Const{\mathop{\rm Const.}\nolimits}
\def\det{\mathop{\rm det}\nolimits}
\def\exp{\mathop{\rm exp}\nolimits}

\def\intr{\mathop{\rm int}\nolimits}
\def\ext{\mathop{\rm ext}\nolimits}

\def\re{\mathop{\rm Re}\nolimits}

\def\lim{\mathop{\rm lim}\nolimits}

\def\sup{\mathop{\rm sup}\nolimits}
\def\inf{\mathop{\rm inf}\nolimits}

\def\Jac{\mathop{\rm Jac}\nolimits}

\def\Sum{\displaystyle\sum}

\def\min{\mathop{\rm min}\nolimits}
\def\max{\mathop{\rm max}\nolimits}

\baselineskip 15pt




\edef\resetatcatcode{\catcode`\noexpand\@\the\catcode`\@\relax}
\ifx\miniltx\undefined\else \fi
\let\miniltx\box

\def\makeatletter{\catcode`\@11\relax}

\makeatletter

\def\@makeother#1{\catcode`#1=12\relax}

\def\@ifnextchar#1#2#3{%
  \let\reserved@d=#1%
  \def\reserved@a{#2}\def\reserved@b{#3}%
  \futurelet\@let@token\@ifnch}
\def\@ifnch{%
  \ifx\@let@token\@sptoken
    \let\reserved@c\@xifnch
  \else
    \ifx\@let@token\reserved@d
      \let\reserved@c\reserved@a
    \else
      \let\reserved@c\reserved@b
    \fi
  \fi
  \reserved@c}
\begingroup
\def\:{\global\let\@sptoken= } \:  
\def\:{\@xifnch} \expandafter\gdef\: {\futurelet\@let@token\@ifnch}
\endgroup

\def\@ifstar#1{\@ifnextchar *{\@firstoftwo{#1}}}
\long\def\@dblarg#1{\@ifnextchar[{#1}{\@xdblarg{#1}}}
\long\def\@xdblarg#1#2{#1[{#2}]{#2}}

\long\def \@gobble #1{}
\long\def \@gobbletwo #1#2{}
\long\def \@gobblefour #1#2#3#4{}
\long\def\@firstofone#1{#1}
\long\def\@firstoftwo#1#2{#1}
\long\def\@secondoftwo#1#2{#2}

\def\NeedsTeXFormat#1{\@ifnextchar[\@needsf@rmat\relax}
\def\@needsf@rmat[#1]{}
\def\ProvidesPackage#1{\@ifnextchar[%
    {\@pr@videpackage{#1}}{\@pr@videpackage#1[]}}
\def\@pr@videpackage#1[#2]{\wlog{#1: #2}}

\let\DeclareOption\@gobbletwo

\def\RequirePackage{%
  \@fileswithoptions\@pkgextension}
\def\@fileswithoptions#1{%
  \@ifnextchar[
    {\@fileswith@ptions#1}%
    {\@fileswith@ptions#1[]}}
\def\@fileswith@ptions#1[#2]#3{%
  \@ifnextchar[
  {\@fileswith@pti@ns#1[#2]#3}%
  {\@fileswith@pti@ns#1[#2]#3[]}}

\def\@fileswith@pti@ns#1[#2]#3[#4]{%
    \def\reserved@b##1,{%
      \ifx\@nil##1\relax\else
        \ifx\relax##1\relax\else
         \noexpand\@onefilewithoptions##1[#2][#4]\noexpand\@pkgextension
        \fi
        \expandafter\reserved@b
      \fi}%
      \edef\reserved@a{\zap@space#3 \@empty}%
      \edef\reserved@a{\expandafter\reserved@b\reserved@a,\@nil,}%
  \reserved@a}

\def\zap@space#1 #2{%
  #1%
  \ifx#2\@empty\else\expandafter\zap@space\fi
  #2}

\let\@empty\empty
\def\@pkgextension{sty}

\def\@onefilewithoptions#1[#2][#3]#4{%
  \input #1.#4 }

\def\typein{%
  \let\@typein\relax
  \@testopt\@xtypein\@typein}
\def\@xtypein[#1]#2{%
  \message{#2}%
  \advance\endlinechar\@M
  \read\@inputcheck to#1%
  \advance\endlinechar-\@M
  \@typein}
\def\@namedef#1{\expandafter\def\csname #1\endcsname}
\def\@nameuse#1{\csname #1\endcsname}
\def\@cons#1#2{\begingroup\let\@elt\relax\xdef#1{#1\@elt #2}\endgroup}
\def\@car#1#2\@nil{#1}
\def\@cdr#1#2\@nil{#2}
\def\@carcube#1#2#3#4\@nil{#1#2#3}
\def\@preamblecmds{}

\def\@star@or@long#1{%
  \@ifstar
   {\let\l@ngrel@x\relax#1}%
   {\let\l@ngrel@x\long#1}}

\let\l@ngrel@x\relax
\def\newcommand{\@star@or@long\new@command}
\def\new@command#1{%
  \@testopt{\@newcommand#1}0}
\def\@newcommand#1[#2]{%
  \@ifnextchar [{\@xargdef#1[#2]}%
                {\@argdef#1[#2]}}
\long\def\@argdef#1[#2]#3{%
   \@ifdefinable #1{\@yargdef#1\@ne{#2}{#3}}}
\long\def\@xargdef#1[#2][#3]#4{%
  \@ifdefinable#1{%
     \expandafter\def\expandafter#1\expandafter{%
          \expandafter
          \@protected@testopt
          \expandafter
          #1%
          \csname\string#1\expandafter\endcsname
          {#3}}%
       \expandafter\@yargdef
          \csname\string#1\endcsname
           \tw@
           {#2}%
           {#4}}}
\def\@testopt#1#2{%
  \@ifnextchar[{#1}{#1[#2]}}
\def\@protected@testopt#1{
  \ifx\protect\@typeset@protect
    \expandafter\@testopt
  \else
    \@x@protect#1%
  \fi}
\long\def\@yargdef#1#2#3{%
  \@tempcnta#3\relax
  \advance \@tempcnta \@ne
  \let\@hash@\relax
  \edef\reserved@a{\ifx#2\tw@ [\@hash@1]\fi}%
  \@tempcntb #2%
  \@whilenum\@tempcntb <\@tempcnta
     \do{%
         \edef\reserved@a{\reserved@a\@hash@\the\@tempcntb}%
         \advance\@tempcntb \@ne}%
  \let\@hash@##%
  \l@ngrel@x\expandafter\def\expandafter#1\reserved@a}
\long\def\@reargdef#1[#2]#3{%
  \@yargdef#1\@ne{#2}{#3}}
\def\renewcommand{\@star@or@long\renew@command}
\def\renew@command#1{%
  {\escapechar\m@ne\xdef\@gtempa{{\string#1}}}%
  \expandafter\@ifundefined\@gtempa
     {\@latex@error{\string#1 undefined}\@ehc}%
     {}%
  \let\@ifdefinable\@rc@ifdefinable
  \new@command#1}
\long\def\@ifdefinable #1#2{%
      \edef\reserved@a{\expandafter\@gobble\string #1}%
     \@ifundefined\reserved@a
         {\edef\reserved@b{\expandafter\@carcube \reserved@a xxx\@nil}%
          \ifx \reserved@b\@qend \@notdefinable\else
            \ifx \reserved@a\@qrelax \@notdefinable\else
              #2%
            \fi
          \fi}%
         \@notdefinable}
\let\@@ifdefinable\@ifdefinable
\long\def\@rc@ifdefinable#1#2{%
  \let\@ifdefinable\@@ifdefinable
  #2}
\def\newenvironment{\@star@or@long\new@environment}
\def\new@environment#1{%
  \@testopt{\@newenva#1}0}
\def\@newenva#1[#2]{%
   \@ifnextchar [{\@newenvb#1[#2]}{\@newenv{#1}{[#2]}}}
\def\@newenvb#1[#2][#3]{\@newenv{#1}{[#2][#3]}}
\def\renewenvironment{\@star@or@long\renew@environment}
\def\renew@environment#1{%
  \@ifundefined{#1}%
     {\@latex@error{Environment #1 undefined}\@ehc
     }{}%
  \expandafter\let\csname#1\endcsname\relax
  \expandafter\let\csname end#1\endcsname\relax
  \new@environment{#1}}
\long\def\@newenv#1#2#3#4{%
  \@ifundefined{#1}%
    {\expandafter\let\csname#1\expandafter\endcsname
                         \csname end#1\endcsname}%
    \relax
  \expandafter\new@command
     \csname #1\endcsname#2{#3}%
     \l@ngrel@x\expandafter\def\csname end#1\endcsname{#4}}

\def\providecommand{\@star@or@long\provide@command}
\def\provide@command#1{%
  {\escapechar\m@ne\xdef\@gtempa{{\string#1}}}%
  \expandafter\@ifundefined\@gtempa
    {\def\reserved@a{\new@command#1}}%
    {\def\reserved@a{\renew@command\reserved@a}}%
   \reserved@a}%

\def\@ifundefined#1{%
  \expandafter\ifx\csname#1\endcsname\relax
    \expandafter\@firstoftwo
  \else
    \expandafter\@secondoftwo
  \fi}

\chardef\@xxxii=32
\mathchardef\@Mi=10001
\mathchardef\@Mii=10002
\mathchardef\@Miii=10003
\mathchardef\@Miv=10004

\newcount\@tempcnta
\newcount\@tempcntb
\newif\if@tempswa\@tempswatrue
\newdimen\@tempdima
\newdimen\@tempdimb
\newdimen\@tempdimc
\newbox\@tempboxa
\newskip\@tempskipa
\newskip\@tempskipb
\newtoks\@temptokena

\long\def\@whilenum#1\do #2{\ifnum #1\relax #2\relax\@iwhilenum{#1\relax
     #2\relax}\fi}
\long\def\@iwhilenum#1{\ifnum #1\expandafter\@iwhilenum
         \else\expandafter\@gobble\fi{#1}}
\long\def\@whiledim#1\do #2{\ifdim #1\relax#2\@iwhiledim{#1\relax#2}\fi}
\long\def\@iwhiledim#1{\ifdim #1\expandafter\@iwhiledim
        \else\expandafter\@gobble\fi{#1}}
\long\def\@whilesw#1\fi#2{#1#2\@iwhilesw{#1#2}\fi\fi}
\long\def\@iwhilesw#1\fi{#1\expandafter\@iwhilesw
         \else\@gobbletwo\fi{#1}\fi}
\def\@nnil{\@nil}
\def\@empty{}
\def\@fornoop#1\@@#2#3{}
\long\def\@for#1:=#2\do#3{%
  \expandafter\def\expandafter\@fortmp\expandafter{#2}%
  \ifx\@fortmp\@empty \else
    \expandafter\@forloop#2,\@nil,\@nil\@@#1{#3}\fi}
\long\def\@forloop#1,#2,#3\@@#4#5{\def#4{#1}\ifx #4\@nnil \else
       #5\def#4{#2}\ifx #4\@nnil \else#5\@iforloop #3\@@#4{#5}\fi\fi}
\long\def\@iforloop#1,#2\@@#3#4{\def#3{#1}\ifx #3\@nnil
       \expandafter\@fornoop \else
      #4\relax\expandafter\@iforloop\fi#2\@@#3{#4}}
\def\@tfor#1:={\@tf@r#1 }
\long\def\@tf@r#1#2\do#3{\def\@fortmp{#2}\ifx\@fortmp\space\else
    \@tforloop#2\@nil\@nil\@@#1{#3}\fi}
\long\def\@tforloop#1#2\@@#3#4{\def#3{#1}\ifx #3\@nnil
       \expandafter\@fornoop \else
      #4\relax\expandafter\@tforloop\fi#2\@@#3{#4}}
\long\def\@break@tfor#1\@@#2#3{\fi\fi}
\def\@removeelement#1#2#3{%
  \def\reserved@a##1,#1,##2\reserved@a{##1,##2\reserved@b}%
  \def\reserved@b##1,\reserved@b##2\reserved@b{%
    \ifx,##1\@empty\else##1\fi}%
  \edef#3{%
    \expandafter\reserved@b\reserved@a,#2,\reserved@b,#1,\reserved@a}}

\let\ExecuteOptions\@gobble

\def\@latex@error#1#2{%
  \errhelp{#2}\errmessage{#1}}

\bgroup\uccode`\!`\%\uppercase{\egroup
\def\@percentchar{!}}

\ifx\@@input\@undefined
 \let\@@input\input
\fi

\def\input{\@ifnextchar\bgroup\@iinput\@@input}
\def\@iinput#1{\input{#1} }

    \def\filename@parse#1{%
      \let\filename@area\@empty
      \expandafter\filename@simple#1.\\}

  \def\filename@simple#1.#2\\{%
    \ifx\\#2\\%
       \let\filename@ext\relax
    \else
       \edef\filename@ext{\filename@dot#2\\}%
    \fi
    \edef\filename@base{#1}}
  \def\filename@dot#1.\\{#1}

\long\def \IfFileExists#1#2#3{%
  \openin\@inputcheck#1 %
  \ifeof\@inputcheck
    \ifx\input@path\@undefined
      \def\reserved@a{#3}%
    \else
      \def\reserved@a{\@iffileonpath{#1}{#2}{#3}}%
    \fi
  \else
    \closein\@inputcheck
    \edef\@filef@und{#1 }%
    \def\reserved@a{#2}%
  \fi
  \reserved@a}
\long\def\@iffileonpath#1{%
  \let\reserved@a\@secondoftwo
  \expandafter\@tfor\expandafter\reserved@b\expandafter
             :\expandafter=\input@path\do{%
    \openin\@inputcheck\reserved@b#1 %
    \ifeof\@inputcheck\else
      \edef\@filef@und{\reserved@b#1 }%
      \let\reserved@a\@firstoftwo%
      \closein\@inputcheck
      \@break@tfor
    \fi}%
  \reserved@a}
\long\def \InputIfFileExists#1#2{%
  \IfFileExists{#1}%
    {#2\@addtofilelist{#1}\@@input \@filef@und}}

\chardef\@inputcheck0

\let\@addtofilelist \@gobble

\def\@defaultunits{\afterassignment\remove@to@nnil}
\def\remove@to@nnil#1\@nnil{}

\newdimen\leftmarginv
\newdimen\leftmarginvi

\newdimen\@ovxx
\newdimen\@ovyy
\newdimen\@ovdx
\newdimen\@ovdy
\newdimen\@ovro
\newdimen\@ovri
\newdimen\@xdim
\newdimen\@ydim
\newdimen\@linelen
\newdimen\@dashdim

\long\def\mbox#1{\leavevmode\hbox{#1}}

\let\@onlypreamble\@gobble

\let\protect\relax

\newdimen\fboxsep
\newdimen\fboxrule

\fboxsep = 3pt
\fboxrule = .4pt

\def\@height{height} \def\@depth{depth} \def\@width{width}
\def\@minus{minus}
\def\@plus{plus}
\def\hb@xt@{\hbox to}

\long\def\@begin@tempboxa#1#2{%
   \begingroup
     \setbox\@tempboxa#1{\color@begingroup#2\color@endgroup}%
     \def\width{\wd\@tempboxa}%
     \def\height{\ht\@tempboxa}%
     \def\depth{\dp\@tempboxa}%
     \let\totalheight\@ovri
     \totalheight\height
     \advance\totalheight\depth}
\let\@end@tempboxa\endgroup

\let\set@color\relax
\let\color@begingroup\relax
\let\color@endgroup\relax
\let\color@setgroup\relax

\let\color@hbox\relax
\let\color@vbox\relax
\let\color@endbox\relax


\begingroup
  \catcode`P=12
  \catcode`T=12
  \lowercase{
    \def\x{\def\rem@pt##1.##2PT{##1\ifnum##2>\z@.##2\fi}}}
  \expandafter\endgroup\x
\def\strip@pt{\expandafter\rem@pt\the}


\def\@input#1{%
  \IfFileExists{#1}{\@@input\@filef@und}{\message{No file #1.}}}

\def\@warning{\immediate\write16}

\def\Gin@driver{dvips.def}
\input graphicx.sty

\resetatcatcode

\medskip
\medskip

\centerline{\bf {VORTICES AND MAGNETIZATION IN KAC'S MODEL}}
\medskip

\centerline{H.EL BOUANANI and M.ROULEUX}
\medskip
\medskip

\centerline{Centre de Physique Th\'eorique and Universit\'e du Sud
Toulon Var}

\centerline{CPT, Campus de Luminy, Case 907 13288 Marseille cedex
9, France}

\centerline{\it{hicham.el-bouanani@cpt.univ-mrs.fr \&
rouleux@cpt.univ-mrs.fr}}
\vskip 2truecm
{\bf Abstract}. We consider a 2-dimensional planar rotator on a
large, but finite lattice with a ferromagnetic Kac potential
$J_\gamma(i)=\gamma^2J(\gamma i)$, $J$ with compact support. The
system is subject to boundary conditions with vorticity. Using a
gradient-flow dynamics, we compute minimizers of the free energy
functional at low temperature, i.e. in the regime of phase
transition. We have the numerical evidence of a vortex structure
for minimizers,  which present many common features with those of
the Ginzburg-Landau functional. We extend the results to spins
valued in $S^2$ and compare with the celebrated  Belavin \& Polyakov
model.

\medskip
\noindent {\bf 0. Introduction}.
\medskip
Vector spin models with an internal continuous symmetry group,
such classical $O^+(q)$ models (XY or ``planar  rotator'' for
$q=2$, and Heisenberg model for $q=3$,~)  play an important r\^ole
in Statistical Physics. In one or two dimensions, and for all
inverse temperature $\beta$, if the range of the translation
invariant interaction is finite, then a theorem of Dobrushin \&
Shlosman shows there is no breaking of the internal symmetry (that
is, Gibbs states are invariant under $O^+(q)$) and furthermore, by
a theorem of Bricmont, Fontaine \& Landau, uniqueness of the Gibbs
state holds (see e.g. [Si,Chap.III]).

Despite
of this, a particular form for phase transition exists, which can
be characterized by the change of behavior in the correlation
functions. In the low temperature phase they have power law decay,
showing that the system is in a long range order state (exhibiting
in particular the so-called ``spin waves'',~) but they decay
exponentially fast at high temperatures, breaking the long range
order, even though thermodynamic quantities remain smooth across
the transition. For the XY system, these transitions were
described by Kosterlitz \& Thouless in term of topological
excitations called vortices~: while these vortices are organized
into dipoles at low temperature, a disordered state emerges at the
transition. But the observation of the spatial distribution of
defects shows that it is not uniform~; rather, defects tend to
cluster at temperatures slightly larger than the transition
temperature, and there are still large ordered domains where the
spins are almost parallel (see e.g. [LeVeRu], [BuPi],[MiZh], and
references therein.~)

Here we consider a Kac version of the classical XY or Heisenberg
model on a ``large'' lattice $\Lambda\subset{\bf Z}^2$.
It was studied in particular by Butt\`a \& Picco [BuPi].
The hamiltonian (except for the interaction with the boundary) is of the form
$$H_\gamma(\sigma_\Lambda)=-{1\over 2}\Sum_{i,j\in\Lambda}\gamma^2 J(\gamma(i-j))
\langle\sigma_\Lambda(i),\sigma_\Lambda(j) \rangle$$ where
$\gamma$ is a small coupling constant and $J$ denotes a cutoff
function. Kac potentials for fixed $\gamma$  have finite interaction,
but as we take an appropriate limit $\gamma\to0$,
they can be considered, to this respect, as
long range. Thus, they share
some features with the mean field model, though exhibiting better
mechanisms of phase transitions, which depend in particular on the
dimension, as for the short range case. For the mean field model
with $O^+(q)$ symmetry, $q=2,3$, we know that there is no phase
transition for inverse temperature $\beta\leq q$ (Gibbs measure is
supported at the absolute minimum of the free energy functional,~)
while there is a phase transition for $\beta>q$, with internal
symmetry group $O^+(q)$.

When the model possesses internal symmetry and common features
with the mean field, it is hard to expect vortices at low
temperature, unless the symmetry is somehow broken, for instance
if the system is subject to boundary conditions. This situation is
met in other domains of Condensed Matter Physics, as in
supraconductivity, where vorticity is created  by an exterior
magnetic flux, or for superfluids. In that case, phase transitions
of matter are well described by critical points of free energy
(Ginzburg-Landau) functionals ([BeBrHe], [OvSi], etc\dots )

One of the main process consists in
averaging the spins $\sigma_\Lambda$ over some mesoscopic boxes, so to define the magnetization
$m=m_{\Lambda^*}$ on another ``coarser'' or ``mesoscopic'' lattice
$\Lambda^*$.
The free energy (or excess free energy) functional
$F_{\beta,\gamma}(m)$ at inverse temperature $\beta$
in case of Kac models with internal symmetry, can be simply
derived from a suitable renormalization of $H_\gamma$  making use
of the entropy $I(m)$ for the mean field  that corresponds to Van der Waals free energy
$f_\beta(m)=-{1\over 2}|m|^2+{1\over\beta}I(m)$ (see Sect.1).

To understand the significance of $F_{\beta,\gamma}(m_{\Lambda^*})$,
one should think also of the formal ``stationary phase'' argument,
as $\Lambda\to\infty$, which suggests that an important r\^ole in
the averaging with respect to Gibbs measure, is played by
configurations close to those which produce the local critical
points of $F_{\beta,\gamma}$. This occurs in computing
correlations functions (see e.g. [Z].~) These critical points
consist in ground states, or metastable states.

They will be determined as the attractors of a certain dynamics,
similar to this given by the ``heat operator'', but known in that
context as the gradient-flow dynamics [DeMOrPrTr], [DeM], [Pr] \dots.
Thus, we expect convergence of this dynamics toward a Gibbsian
equilibrium, though this will not be rigorously established here.

Let us present our main results.

In Sect.1, we describe in detail Kac's hamiltonian on the lattice,
and recall briefly the renormalization scheme, that makes of
the free energy functional a fairly good approximation for the density of
Gibbs measure, i.e.
$\mu_{\beta,\gamma,\Lambda}\approx\exp[-\beta\gamma^{-2}{\cal F}(m|m^c)]$.
Here ${\cal F}(m|m^c)$ denotes the free energy functional subject to boundary conditions $m^c$ on $\Lambda^{*c}$.

In Sect.2, we present a simple, combinatorial averaging process,
relating Kac's hamiltonian with
the free energy functional. While exhibiting the main idea of renormalization, it is more
suitable for effective computations on the lattice.

In Sect.3, we study Euler-Lagrange equations for the free energy functional,
and introduce the corresponding
gradient-flow dynamics $m(x,t)$ (see Eqn. (3.4).~) Using that ${\cal F}(m|m^c)$
is a Lyapunov function, we show
that $m(x,t)$ converges towards a critical point of ${\cal F}(m|m^c)$, generically, a local minimum.
Unless $\beta\leq2$, in which case $m=0$
is the unique minimizer of ${\cal F}(m|m^c)$, as expected from the considerations above on the mean field,
in general there cannot be uniqueness of the limiting orbits, at least for a finite lattice.
Instead, local minimizers might depend on initial conditions $m(0,x)$ inside $\Lambda^*$.

Local minimizers however, have the property that their modulus be bounded by $m_\beta$, if this is
true of the initial condition,
and as expected from general results relative to the Gibbs states [BuPi],
$|m|$ has to be close to $m_\beta$ on large regions of $\Lambda^*$.
Actually Proposition 3.4 indicates that if no vorticity is induced by the boundary, nor by the initial condition,
then all magnetizations of
the limiting configuration should point out in the same direction and have length about $m_\beta$.

In Sect.4 we make numerical simulations, introducing
a boundary condition with topological degree $d\in{\bf Z}$.
Then, on the basis on conservation of vorticity, the limiting orbits for the gradient-flow dynamics show a vortex pattern.
For $q=2$, our main observation is the existence of vortices below
the temperature of transition of phase for the mean field model,
induced by the vorticity at the boundary of the lattice $\Lambda$,
together with large ordered domains where the magnetizations
$m_{\Lambda^*}$ become parallel. We discuss in detail dependence on the shape of the lattice, and on initial conditions.
In particular, the application of the``simulated annealing process'' allows the limiting configurations to move away from
local minima, and reach lower energies.

We also have some numerical evidence that, as in the case of
Ginzburg-Landau functional, Kirchhoff-Onsager hamiltonian for the
system of vortices gives a fairly good approximation of the
minimizing free energy, despite of the non-local interactions.

Finally, for $q=3$, we examine in Sect.5 the situation of spin-waves in the
spirit of Belavin \& Polyakov.

\medskip
\noindent {\bf Acknowledgements}:  We are very grateful to P.
Picco who introduced us to the subject~; we also thank A. Messager and Y.Vignaud for many
interesting and useful discussions.

\medskip
\smallskip
\noindent  {\bf 1. Mean field approximation and renormalized Kac's
Hamiltonian}.  
\smallskip
Consider the lattice ${\bf Z}^2$, consisting in a bounded,
connected domain $\Lambda$ (the interior region), and its
complement (the exterior region) $\Lambda^c$. In practice, we
think of $\Lambda$ as a large rectangle with sides parallel to the
axis of ${\bf  Z}^2$, of length of the form $L=2^n$, $n\in{\bf
N}$. Physical objects make sense in the thermodynamical limit
$\Lambda\to{\bf Z}^2$, but in this paper we work in large, but
finite domains.

To each site $i\in{\bf Z}^2$ is attached a classical spin variable
$\sigma_i\in{\bf S}^{q-1}$, $q=2,3$. The configuration space
${\cal X}({\bf Z}^2)=({\bf S}^{q-1})^{{\bf Z}^2}$ is the set of
all such classical states of spin~; it has the natural internal
symmetry group $O^+(q)$ acting on ${\bf S}^{q-1}$. The state
$\sigma\in {\cal X}({\bf Z}^2)$ will denote the map $\sigma:{\bf
Z}^2\to{\bf S}^{q-1}$, $i\mapsto\sigma(i)$. Given the partition
${\bf Z}^2=\Lambda\cup\Lambda^c$, we define by restriction the
interior and exterior configuration spaces ${\cal X}(\Lambda)$ and
${\cal X}(\Lambda^c)$, and the restricted  configurations by
$\sigma_\Lambda$ and $\sigma_{\Lambda^c}$. The Hamiltonian in
${\bf Z}^2$ describes the interaction between different sites
through Kac's potential defined as follows.

Let $0\leq J\leq 1$ be a function on ${\bf R}^2$ with compact
support and normalized by $\int_{{\bf R}^2}J=1$. We can think of
$J$ also as a function on the lattice. There is a lot of freedom
concerning the choice of $J$, but for numerical purposes, we take
$J$ as 1/2 the indicator function $\widetilde J$ of the unit
rhombus with center at the origin, in other words $J(x)=J(|x|_1)$
where $|\cdot|_1$ is the $\ell^1$ norm in ${\bf R}^2$. Thus the
support of $\widetilde J$ is thought of as a chip of area 2, and
considered as a function on the lattice, $\widetilde J$ takes the
value 1 at the center, and 1/4 at each vertex, so that
$\Sum_{i\in\Lambda}J(i)=1$. For $\gamma$ of the form $2^{-m}$, we
set $J_\gamma(x)=\gamma^2J(\gamma x)$, and extend the definition
above in the discrete case so that $J_\gamma$ enjoys good scaling
properties, namely the stratum of full dimension (i.e. the set of
points interior to the chip) has weight 1, the strata of dimension
1 (the points on the sides on the chip) have weight 1/2, and those
of dimension 0 (the vertices of the chip) have weight 1/4. Thus,
again $\Sum_{i\in{\bf Z}^2}J_\gamma(i)=1$. The discrete
convolution on $\Lambda$ is defined as usual. For instance,
$(J_\gamma*\sigma)(i)=\Sum_{j\in{\bf Z}^2}J_\gamma(i-j)\sigma(j)$
represents, with conventions as above, the mean value of $\sigma$
over the chip of size $\gamma^{-1}$ and center $i$, with a weight
that depends on the stratum containing $j$.

Note that we could replace the lattice ${\bf Z}^2$ by the torus
$({\bf Z}/L{\bf Z})^2$ or the cylinder $({\bf Z}/L{\bf Z})\times
{\bf Z}$, which amounts to specify periodic boundary conditions in
one or both directions. Thermodynamic limit is obtained as
$L\to\infty$.

The coupling  between spin at site $i$ and  spin at site $j$ is
given by $J_\gamma(i-j)$~; this is known as Kac's potential. From
Statistical Physics point of view, Kac's potential,  for small
$\gamma$, shares locally the main properties of the mean field,
i.e. long range $\approx \gamma^{-1}$, large connectivity
$\approx\gamma^{-2}$ of each site, small coupling constant
$\approx\gamma^2$ of the bonds, and total strength of each site
equal to 1.

Given the exterior configuration $\sigma_{\Lambda^{c}}\in{\cal
X}(\Lambda^{c})$, we define the Hamiltonian on ${\bf Z}^2$ as
$$H_\gamma(\sigma_{\Lambda}|\sigma_{\Lambda^c})
=-{1\over 2}\Sum_{i,j\in\Lambda}J_\gamma(i-j)\langle\sigma(i),
\sigma(j)\rangle- \Sum_{(i,j)\in\Lambda\times\Lambda^c}
J_\gamma(i-j)\langle\sigma(i), \sigma(j)\rangle \leqno(1.1)$$
where $\sigma(i)$, for simplicity, stands for
$\sigma_{\Lambda}(i)$ or $\sigma_{\Lambda^c}(i)$, and
$\langle\cdot,\cdot\rangle$ is the standard scalar product in
${\bf R}^q$. We note that as $J\geq 0$, the interaction  is
ferromagnetic, i.e. energy decreases as spins align.

We give here some heuristic derivation of the model we will
consider, starting from principles of Statistical Physics. A
thermodynamical system at equilibrium is described by Gibbs
measure at inverse temperature $\beta$. We assume an a priori
probability distribution $\nu$ for the states of spin, and because
of the internal continuous symmetry of ${\cal X}(\Lambda)$, we
take $\nu$ as the normalized surface measure on ${\bf S}^{q-1}$,
i.e. $\nu(d\sigma_i)=\omega_q^{-1}\delta(|\sigma_i|-1)d\sigma_i$,
where $\omega_q$ is the volume of ${\bf S}^{q-1}$. Then Gibbs
measure on ${\cal X}(\Lambda)$ with prescribed boundary condition
$\sigma_{\Lambda^c}$ is given by
$$\mu_{\beta,\gamma}(d\sigma_{\Lambda}|
\sigma_{\Lambda^c}) ={1\over
Z_{\beta,\gamma}^\Lambda(\sigma_{\Lambda^c})} \exp \bigl[-\beta
H_\gamma(\sigma_{\Lambda}|\sigma_{\Lambda^c})
\bigr]\prod_{i\in\Lambda}\nu\bigl(d\sigma_{\Lambda}(i)\bigr)\leqno(1.2)
$$
where $Z^{\Lambda}_{\beta,\gamma}(\sigma_{\Lambda^c})$, the
partition function, is a normalization factor which makes of
$\mu_{\beta,\gamma}$ a probability measure on ${\cal X}(\Lambda)$,
conditioned by $\sigma_{\Lambda^c}\in{\cal X}(\Lambda^c)$. It is
obtained by integration of $\mu_{\beta,\gamma}(d\sigma_{\Lambda}|
\sigma_{\Lambda^c})$ over $\Omega_0=\bigl({\bf S}^{q-1}\bigr)^\Lambda$.

Since we are working on ${\bf Z}^2$, there exists, for any
$\beta>0,\gamma>0$, an infinite volume Gibbs state
$\mu_{\beta,\gamma}$, i.e. a (unique) probability distribution
$\mu_{\beta,\gamma}$ on the space ${\cal X}$ of all configurations
obtained by taking the thermodynamic limit $\Lambda\to{\bf Z}^2$.
This measure satisfies suitable coherence conditions, i.e. DLR
equations.

Nevertheless, we are faced with various difficulties, indicating
that $\mu_{\beta,\gamma}(d\sigma_{\Lambda}|
\sigma_{\Lambda^c})$ should not be the object to be directly considered. It is known that
(and this goes back to Van Hove for the Ising ferromagnet, i.e. $q=1$, see [Si,p.31],~)
in order to understand thermodynamical
properties for spins models, one should instead average spins over
mesoscopic regions and consider the image of Gibbs measure through
this transformation,  the so called ``block-spin transformation''.
So we introduce the empirical magnetization in the
finite box $\Delta\subset{\bf Z}^2$
$$m_{\Delta}(\sigma)={1\over|\Delta|}\Sum_{i\in\Delta}\sigma(i)\leqno(1.3)$$
and given any $m\in{\bf R}^q$, $|m|\leq 1$, we define the canonical partition function in $\Delta$ as
$$Z_{\beta,\gamma}^{\Delta,\sigma_{\Delta^c}}(m)=
\int_{\bigl({\bf S}^{q-1}\bigr)^\Delta} \exp \bigl[-\beta
H_\gamma(\sigma_{\Delta}|\sigma_{\Delta^c})\bigr]
\prod_{i\in\Delta}\nu\bigl(d\sigma(i)\bigr)\delta\bigl(m_\Delta(\sigma)-m\bigr)
\leqno(1.4)$$
see [Si,p.31]. Also for Kac's model,
it can be shown, taking the thermodynamical limit
$\Delta\to{\bf Z}^2$, that the quantity
$$F_\gamma(\beta,m)=-\lim_{\Delta\to{\bf Z}^2}
{1\over\beta|\Delta|}\log
Z_{\beta,\gamma}^{\Delta,\sigma_{\Delta^c}}(m) \leqno(1.5)$$
is well defined, and doesn't depend on the boundary condition on
$\Delta^c$~; it will be interpreted as the thermodynamic free energy density of
the system. It is defined for a system with finite interaction of range $\gamma^{-1}$, i.e. before taking the mean field limit
$\gamma\to 0$.

So far, parameter $\gamma$ was kept small
but constant~; the limit $\gamma\to 0$ is called Lebowitz-Penrose
limit. Let
$$f_\beta(m)=-{1\over 2}|m|^2+{1\over\beta}I(m)\leqno(1.6)$$
be the free energy for the mean field, $I(m)$ denotes
the entropy, see (2.2) below.

Lebowitz-Penrose theorem (in this simplified context) states that
$$\lim_{\gamma\to 0}F_\gamma(\beta,m)=\hbox{CE} \bigl(f_\beta(m)\bigr)\leqno(1.7)$$
See [BuPi] for the case
of a 1-d lattice and continuous symmetry, the proof can be carried over to ${\bf Z}^2$. Here CE
denotes the convex envelope, to account for Maxwell correction law.

From this we sketch the renormalization procedure that leads to
Lebowitz-Penrose theorem, as stated e.g. in [Pr,Thm. 3.2.1] for $q=1$, following earlier results by [AlBeCaPr]
(actually, this is the ``pressure'' version of Lebowitz-Penrose
theorem, but the argument can easily be adapted to free energy.~)
The following result will not be used in the sequel, we just give it for completeness.

Since we will take (in this paragraph) a continuous limit, we do
assume  that $J$ is a differentiable function, not necessarily of
compact support, but with $\|\nabla J\|_1<\infty$ (the $L^1$
norm.~) The lattice dimension $d$ can be arbitrary. We set,
following (1.3), $\Delta=\widetilde\Lambda(x)$ and
$m_\sigma(x)=m_{\widetilde \Lambda(x)}(\sigma)$. Here $\widetilde
\Lambda(x)$ will be a square ``centered'' at a variable $x\in{\bf
Z}^2$, with sides of length ${\delta\over\gamma}$, $\delta$ of the
form $2^{-p}$, $p\in{\bf N}$, ${\delta\over\gamma}$ much smaller
than the diameter of $\Lambda$, but still containing many sites,
for instance diam $(\widetilde\Lambda(x))=\gamma^{-1/2}$.
Actually, we need to replace (1.3) by an integral, which allows to
extend $x\mapsto m_\sigma(x)$ on ${\bf R}^d$, but for simplicity,
we present it as a discrete sum.

The averages $m_\sigma(x)$ are called (empirical) magnetizations.
The set of all such magnetizations
$m_\sigma\in{\bf R}^q$ is the image of ${\cal X}({\bf Z}^2)$ by
the block-spin transformation $\pi_\gamma:\sigma\to m_\sigma$, and will be denoted by
$\widetilde {\cal X}({\bf Z}^2)$. This is the set of ``coarsed-grained''
configurations.

It has again the continuous
symmetry group $O^+(q)$, and this is a subset of the convex set
${\cal M}$ of all functions $m:{\bf Z}^2\to{\bf R}^q$ such that
$|m(x)|\leq 1$ for all $x$. When considering microscopic
interior and exterior regions as above, the partition ${\bf
Z}^2=\Lambda\cup\Lambda^c$ induces of course restricted
configuration spaces $\widetilde{\cal X}(\Lambda^*)$ and
$\widetilde{\cal X}(\Lambda^{*c})$, where $\Lambda^*=\{x\in{\bf
Z}^2: \widetilde\Lambda(x)\subset\Lambda\}$ and
$\Lambda^{*c}=\{x\in{\bf Z}^2:
\widetilde\Lambda(x)\subset\Lambda^c\}$. So let $m\in{\cal M}$.

We introduce as in (1.4) the canonical Gibbs measure conditioned by the external configuration
$\sigma_{\Lambda^c}=\sigma^c$~:
$$\mu_{\beta,\gamma,\Lambda}(d\sigma;m|\sigma^c)=
{1\over  Z_{\beta,\gamma,\Lambda}(\sigma^c)}\int_{\Omega_0}
\prod_{i\in\Lambda}\nu\bigl(d\sigma(i)\bigr)\exp \bigl[-\beta H_\gamma(\sigma(i)|\sigma^c)\bigr]
\delta(\pi_\gamma\sigma(i)-m)\leqno(1.8)$$
where the  partition function
$Z_{\beta,\gamma,\widetilde\Lambda}(\sigma_{\widetilde\Lambda^c})$
was defined  in (1.2). For simplicity, we have removed the index $\Lambda$ from $\sigma$.
By definition
of the image of Gibbs measure through the block-spin transformation, we have
$$\int_{|m|<1}dm\mu_{\beta,\gamma,\Lambda}(d\sigma;m|\sigma_\gamma^c)=1\leqno(1.9)$$
where $dm$ is the normalized Lebesgue measure on the product space $\prod_{x\in\Lambda^*}B_q(0,1)$.
($B_q(0,1)$ denotes the unit ball of ${\bf R}^q$.~) Let
$$\eqalign{
{\cal F}(m|m^c)&={1\over 4}\int_{\Lambda_0} dr\int_{\Lambda_0} dr'J(r-r')|m(r)-m(r')|^2\cr
&+{1\over 2}\int_{\Lambda_0} dr\int_{\Lambda_0^c} dr'J(r-r')|m(r)-m(r')|^2+
\int_{\Lambda_0} dr \bigl(f_\beta(m(r))-f_\beta(m_\beta)\bigr)\cr
}\leqno(1.10)$$
be the continuous, free energy in a
box $\Lambda_0\subset{\bf R}^2$ of fixed size $L_0$, rescaled from $\Lambda$ by a factor proportional to $\gamma$.
Here $m_\beta$ is the critical value for the mean field $f_\beta$, see Sect.2.
Assume, as before, that the diameter of all block spins $\widetilde\Lambda(x)$ equals $\gamma^{-1/2}$.
Then we can give  a special meaning to the approximation
$\mu_{\beta,\gamma,\Lambda}\approx\exp[-\beta\gamma^{-d}{\cal F}(m|m^c)]$
(in the logarithmic sense) stated in the Introduction, by
establishing the analogue of [AlBeCaPr,Lemma 3.2] in case of continuous symmetry, improving also [BuPi, Lemma 3.1].
Let $\widehat e$
be any (fixed) unit vector in ${\bf R}^q$, and $\widehat m_\beta$ the constant function on $\Lambda$
equal to $m_\beta\widehat e$, which we extend to be equal to $m^c$ on $\Lambda^c$.
We have the following~:
\medskip
\noindent{\bf Proposition 1.1}: Let $q=2$. With the notations above, there are constants $C_1, C_2>0$ such that for
any coarse-grained configuration $m$ on $\Lambda^*$~:
$$\eqalign{
-g(m)-&(L_0\gamma^{-1})^d\bigl(C_2\beta\sqrt\gamma\|\nabla
J\|_1+C_1\gamma^{d/2}\log \gamma^{-1}\bigr)\cr \leq\log
&[\mu_{\beta,\gamma,\Lambda}(d\sigma;m|\sigma^c)]+\beta\gamma^{-d}{\cal
F}(m|m^c)\cr &\leq g(m)+\beta\gamma^{-d}\inf_{\widehat e\in{\bf
S^1}}{\cal F}(\widehat m_\beta|m^c)
+(L_0\gamma^{-1})^d\bigl(C_2\beta\sqrt\gamma\|\nabla
J\|_1+C_1\gamma^{d/2}\log \gamma^{-1}\bigr)\cr }\leqno(1.11)$$
where $g(m)=\log\prod_{x\in\Lambda^*}\bigl(1-|m(x)|\bigr)^{-1/2}$.
\smallskip
See [El-BoRo] for details.
The divergence of $g(m)$ as $|m|$ gets close to 1 reflects the fact that the entropy density $I(m)$
is singular at $|m|=1$, precisely where the mean field approximation breaks down, see also [BuPi,Theorem 2.2].
So the approximation $\mu_{\beta,\gamma,\Lambda}\approx\exp[-\beta\gamma^{-d}{\cal F}(m|m^c)]$ holds true
when the magnetization stays bounded away from 1, as is the case in most applications.

Having this construction in mind, we shall proceed the other way
around, and make a simple renormalization of $H_\gamma$ (see
Proposition 2.1 below). Actually  our sole purpose is to give a
discrete analogue for the excess free energy functional as in
(1.10), most adapted to numerical experiments on the lattice.
\medskip
\noindent {\bf 2. Renormalized Hamiltonian on the lattice}.
\smallskip
We restrict here to $q=2$, in Sect. 5 we show how these
considerations easily extend to $q=3$. Recall from (1.6) the free energy for the
mean field, $I(m)$ is the entropy function of the a priori measure
$\nu$, which can be computed following [BuPi]. Namely, introduce the
moment generating function
$$\phi(h)=\int_{S^{q-1}}e^{\langle h,\sigma\rangle}d\nu(\sigma)\leqno(2.1)$$
and define $I(m)$ as Legendre transformation
$$I(m)=\widehat I(|m|)=\sup _{h\in{\bf R}^q}\bigl(\langle h,m\rangle-\log
\phi(h)\bigr) \leqno(2.2)$$
We denote by $I_n$ the modified Bessel
function of order $n$. For $q=2$, we have
$\phi(h)=\widehat\phi(|h|)=I_0(|h|)$. Function
$\rho\mapsto\widehat I(\rho)$ is convex, strictly increasing on
[0, 1], $\widehat I(\rho)\sim\rho^2$ as $\rho\to 0$, $\widehat
I(\rho)\sim-{1\over 2}\log(1-\rho)$, as $\rho\to 1$, and these
relations can be differentiated. We have also $\widehat
I'=\bigl((\log\widehat\phi)'\bigr)^{-1}$ and
$(\log\widehat\phi)'(t)= I_1(t)/I_0(t)$, this is of course a real
valued function. The phase transition of mean field type is given
by the critical point of the free energy $f_\beta$, i.e. the
positive root of equation $\beta m_\beta=\widehat I'(m_\beta)$,
which exists iff $\beta>\widehat I''(0)=2$.  So the critical
manifold has again $O^+(2)$ invariance.

Now we specify the choice of mesoscopic boxes
$\widetilde\Lambda(x)$ and construct the excess free energy
functional by the procedure sketched above. When $q=2$, it is
convenient to use the underlying complex structure of ${\cal
X}({\bf Z}^2)$, so we shall write (1.1), with obvious notations,
as
$$H_\gamma(\sigma_\Lambda|\sigma_{\Lambda^c})
=-{1\over 2}\Sum_{i,j\in\Lambda}J_\gamma(i-j)\sigma(i)
\overline{\sigma(j)}-\re \Sum_{(i,j)\in\Lambda\times\Lambda^c}
J_\gamma(i-j)\sigma(i) \overline{\sigma(j)} \leqno(2.3)$$ We
introduce in detail the mesoscopic ensemble averages, or coarse
graining approximation to renormalize $H_\gamma$. Let $\delta>0$
be small, but still much larger than $\gamma$, we take again
$\delta=2^{-m}$, for some $m\in{\bf N}$. We take for
$\widetilde\Lambda(x)$, $x\in{\bf Z}^2$, a square ``centered'' at
$x$, of diameter ${\delta\over\gamma}$, and of the form
$\widetilde\Lambda_\delta(x)=\{i=(i_1,i_2)\in{\bf Z}^2:
i_k\in{\delta\over\gamma}[x_k,x_k+1[\}$ where we define as in
(1.2), $m_\delta(x)=\bigl({\gamma\over\delta}\bigr)^2
\Sum_{i\in\widetilde\Lambda(x)}\sigma(i)$. Thus we magnify by a
factor $\delta/\gamma$ the ``coarse graining'' (or mesoscopic
ensemble) labelled by $x\in{\bf Z}^2$, to the ``smooth graining''
(or microscopic ensemble) labelled by $i\in{\bf Z}^2$. We have~:
\medskip
\noindent {\bf Proposition 2.1}: There is $0<\alpha<{1\over 4}$
such that
$$\bigl({\gamma\over\delta}\bigr)^2
H_\gamma(\sigma_{\Lambda}|\sigma_{\Lambda^c}) +U_{\ext
}(m_\delta)+U_{\intr }(m_\delta)-|\Lambda|f_\beta(m_\beta)= {\cal
F}(m_\delta|m_\delta^c)+|\Lambda|{\cal
O}\bigl(\delta^{2\alpha}\bigr) \leqno(2.4)$$ where
$$\eqalign{
{\cal F}&(m_\delta|m_\delta^c)={1\over 4}\Sum_{x,y\in\Lambda^*}J_\delta(x-y)
|m_\delta(x)-m_\delta(y)|^2\cr
&+{1\over
2}\Sum_{(x,y)\in\Lambda^*\times \Lambda^{*c}}J_\delta(x-y)
|m_\delta(x)-m_\delta(y)|^2+\Sum_{x\in\Lambda^*}
f_\beta(m_\delta(x))-f_\beta(m_\beta)\cr
&U_{\ext }(m_\delta)={1\over
2}\Sum_{(x,y)\in\Lambda^*\times
\Lambda^{*c}}J_\delta(x-y)|m_\delta(y)|^2\cr &U_{\intr
}(m_\delta)={1\over \beta}\Sum_{x\in\Lambda^*} I(m_\delta(x))\cr
}\leqno(2.5)$$
\smallskip
\noindent {\it Proof}: To start with, consider the first term in
(1.1)
$$\eqalign{
&\bigl({\gamma\over\delta}\bigr)^2
\Sum_{i,j\in\Lambda}J_\gamma(i-j)\sigma(i)\overline{\sigma(j)}=
\Sum_{x,y\in\Lambda^*}J_\delta(x-y)m_\delta(x)
\overline{m_\delta(y)}\cr &+\gamma^2\Sum_{x,y\in\Lambda^*}
\Sum_{(i,j)\in\widetilde\Lambda_\delta(x)\times
\widetilde\Lambda_\delta(y)}
\bigl(J(\gamma(i-j))-J(\delta(x-y))\bigr)
\sigma(i)\overline{\sigma(j)}\cr }\leqno(2.6)$$ and denote by
$R(\Lambda^*)$ the second sum in the RHS of (2.6). Let
$C_0=B_1(0,{1\over\delta})$ be the rhombus (or $\ell^1$-ball in
${\bf R}^2$) of center 0 and radius ${1\over\delta}$,
corresponding to the shape of the interaction $J$, and for
$x'\in{\bf Z}^2$, its translate $C_{x'}={1\over\delta}x'+C_0$, we
denote also by $C^*_{x'}\subset\Lambda^*$ the corresponding
lattice obtained from $C_{x'}$ by deleting 2 of its sides, so that
$\Lambda^*=\bigcup_{x'\in{\bf Z}^2}C^*_{x'}$ (disjoint union), and
$\Lambda^*$ is covered by those $C^*_{x'}$ with $x'=(x'_1,x'_2)$,
$x'_j\in\{\pm 1,\cdots, \pm\gamma L\}$. Let also
$E(x,y)=\{(i,j)\in\widetilde\Lambda_\delta(x)\times\widetilde\Lambda_\delta(y):
J(\gamma(i-j))-J(\delta(x-y))\neq 0\}$. We can consider $E(x,y)$
as a symmetric relation $E:\Lambda^*\to\Lambda^*$,
$E(x)=\{y\in\Lambda^*: E(x,y)\neq \emptyset\}$. By translation
invariance of $J$, for any $x'\in{\bf Z}^2$, we have
$|E(x,y)|=|E(x-{1\over\delta}x', y-{1\over\delta}x')|$, so that
$$\Sum_{x,y\in\Lambda^*}|E(x,y)|\leq 4\bigl({\gamma L\over\delta}
\bigr)^2\Sum_{x,y\in C_0^*}|E(x,y)| \leqno(2.7)$$
With the choice
of $\ell^1$ norm, we have $E(x,y)\neq\emptyset$ for all $x,y\in
C_0^*$, and $\max _{x\in C_0^*}|E(x)|=\bigl(1+{1\over 2\delta}
\bigr)^2$, while $\min _{x\in C_0^*}|E(x)|$ is of order unity. In
any case, $|E(x)|$ depends on $x$ and $\delta$, but not on
$\gamma$, and it is easy to see that for some $0<\alpha<{1\over
4}$, $\Sum_{x\in C_0^*}|E(x)|= {\cal O}(\delta^{-2(1-\alpha)})$,
$\delta\to 0$. [Actually, this kind of estimate is well-known, see
e.g. [BlLe] and references therein for related results, and
applies whenever the support of $J$ is a convex set.~]

On the other hand, we have the rough estimate
$|E(x,y)|\leq|\widetilde\Lambda_\delta(x)\times
\widetilde\Lambda_\delta(y)|=\bigl({\delta\over\gamma}\bigr)^4$,
and since $|\sigma(i)|=1$,
$$\eqalign{
|\Sum_{x,y\in C_0^*}
\Sum_{(i,j)\in\widetilde\Lambda_\delta(x)\times
\widetilde\Lambda_\delta(y)}&
\bigl(J(\gamma(i-j))-J(\delta(x-y))\bigr)\sigma(i)\overline{\sigma(j)}|\cr
&\leq\bigl({\delta\over\gamma}\bigr)^4 \Sum_{x\in
C_0^*}|E(x)|=\bigl({\delta\over\gamma}\bigr)^4 {\cal
O}(\delta^{-2(1-\alpha)})\cr }$$
This, together with (2.7), shows
that $R(\Lambda^*)\leq \Const \delta^{2\alpha}L^2$. A similar
argument gives an estimate on the remainder
$R(\Lambda^*|\Lambda^{*c})$ for the second term in (1.1). Once we
have replaced $\bigl({\gamma\over\delta}\bigr)^2
\Sum_{i,j}J_\gamma(i-j)\sigma(i)\overline{\sigma(j)}$ by
$\Sum_{x,y}J_\delta(x-y)m_\delta(x) \overline{m_\delta(y)}$ modulo
$R(\Lambda^*)$ and $R(\Lambda^*|\Lambda^{*c})$, which
verify the estimate given in (2.4), we use the identity
$$-2\re m_\delta(x)\overline{m_\delta(y)}=
|m_\delta(x)-m_\delta(y)|^2-|m_\delta(x)|^2-|m_\delta(y)|^2$$ and
express the ``density'' term ${1\over 2}|m|^2$ in term of the mean
field free energy $f_\beta(m)$ as in (2.1). Summing over $(x,y)$
and making use of the fact that $J_\delta$ is normalized in
$\ell^1({\bf Z}^2)$ eventually gives the Proposition. $\clubsuit$
\medskip
\noindent {\it Remarks}: 1) In homogenization problems, one
usually associates the discrete configuration $\sigma\in{\cal
X}(\Lambda)$ with the function $\sigma_\gamma$ on ${\bf R}^2$
taking the constant value $\sigma(i)$ on the square ``centered''
at $\gamma i$, $i=(i_1,i_2)$, i.e. on $[\gamma i_1,
\gamma(i_1+1)[\times [\gamma i_2, \gamma(i_2+1)[$. Furthermore the
size of the domain $\Lambda$ is normalized, so that taking the
thermodynamic limit $\Lambda\to\infty$ is a problem of convergence
for piecewise constant functions (or discrete measures) in some
suitable functional space. As we have seen in Sect.1, it is convenient to
take a smooth interaction $J$. Thus a version of Proposition 2.1
was obtained in [BuPi] by replacing the discrete average
$m_\delta(x)$ around $x\in\Lambda$ by an integral, or in
[DeMOrPrTr], [DeM], [Pr], \dots by averaging $J_\gamma$ over boxes
of type $C_{x'}$ as above. (For short we refer henceforth to the
review article [Pr]). Since our ultimate purpose here consists in
numerical simulations on a lattice, we chose instead to give a
discrete renormalization for $H_\gamma$.

2) Our renormalized Hamiltonian is now given by ${\cal
F}(m_\delta|m_\delta^c)$, the quantities we have subtracted are
$-U_{\ext }(m_\delta)$, attached to the configuration space ${\cal
X}(\Lambda^c)$, and $-U_{\intr }(m_\delta)$ that can be
interpreted as $\beta^{-1}$ times the entropy of the system in
$\Lambda$. Note we have also included self-energy terms $i=j$ in
the original Hamiltonian. Of course, relevance of this free energy
to Gibbs measure (or rather its image through the block-spin
transformation) after taking the thermodynamic limit, is a rather
subtle question which will not be discussed here, since we content
to finite lattices.
\medskip
\noindent {\bf 3. Euler-Lagrange equations and non local
dynamics}.
\medskip
We are interested in the critical points of ${\cal F}(m_\delta|m_\delta^c)$. Denote as usual resp. by $\partial_m$
and $\overline \partial_m$ the holomorphic and anti-holomorphic
derivatives, we have for $m=m_\delta$ (for short), and any tangent
vector of type (1,0) in the holomorphic sense, $\delta m\in
T_m^{(1,0)}\widetilde{\cal X}({\bf Z}^2)$~:
$$\eqalign{
&\langle\partial_m{\cal F}(m|m^c),\delta m\rangle= {1\over
2}\Sum_{(x,y)\in\Lambda^*\times \Lambda^{*c}}J_\delta(x-y)\bigl(\overline m(x)-\overline m(y)\bigr) \delta m(x)\cr
&+{1\over
2}\Sum_{x,y\in\Lambda^*}J_\delta(x-y) \bigl(\overline
m(x)-\overline m(y)\bigr)\delta m(x)
+\Sum_{x\in\Lambda^*}\bigl(-{1\over 2}\overline
m(x)+{1\over\beta} {\partial I(m)\over\partial m}(x)\bigr)\delta
m(x)\cr }$$
Using again the normalization of $J_\delta$ in
$\ell^1({\bf Z}^2)$, the relation $I(m)=\widehat I(|m|)$, and
setting as before $J_\delta *m(x)=\Sum_{y\in{\bf
Z}^2}J_\delta(x-y)m(y)$, we obtain
$$\langle\partial_m{\cal F}(m|m^c),\delta m\rangle
={1\over 2}\Sum_{x\in\Lambda^*}\bigl( -J_\delta *\overline
m(x)+{1\over\beta} {\widehat I'(|m|)\over|m|}\overline
m(x)\bigr)\delta m(x)\leqno(3.1)$$ Since ${\cal F}$ is real,
this gives Euler-Lagrange equation~:
$$-J_\delta *m(x)+{1\over\beta}
{\widehat I'(|m|)\over|m|}m(x)=0\leqno(3.2)$$ Let $f=(\widehat
I')^{-1}={\widehat\phi'\over\widehat\phi}$ denote the inverse of
the function $\widehat I'$. Thus $f:[0,+\infty[\to[0,1[$ is
strictly concave, $f(0)=0, f'(0)=1/2$, and $f(\rho)\to 1$ as
$\rho\to +\infty$. Since the inverse of $m\mapsto {\widehat
I'(|m|)}{m\over|m|}$ defined on the unit disk is given by
$n\mapsto f(|n|){n\over|n|}$, $n\in{\bf C}$, (3.2) takes the form
$$-m+f(\beta|J_\delta *m|){J_\delta *m\over |J_\delta *m|}=0\leqno(3.3)$$
Following [Pr], to find the critical points minimizing the excess
free energy functional ${\cal F}$ we solve the ``heat equation''
$${dm\over dt}=-m+f(\beta|J_\delta *m|){J_\delta *m\over |J_\delta *m|}
\ \hbox{in} \ \Lambda^* \leqno(3.4)$$
with prescribed (time
independent) boundary condition on $\Lambda^{*c}$, and initial
condition $m_{|\Lambda^*}=m_0$. By Cauchy-Lipschitz theorem,
equation (3.4) has a unique solution, defined for all $t>0$,
valued in $\widetilde{\cal X}(\Lambda^*)$. Monotonicity of ${\cal
F}$ is given in the following~:
\medskip
\noindent {\bf Proposition 3.1}: ${\cal F}$ is a Lyapunov function
for equation (3.4), i.e. there exists a free energy dissipation rate function
${\cal I}:\widetilde{\cal X}(\Lambda^*)\to{\bf R}^+$, ${\cal I}(m)=0$ iff $m$ solves (3.3), and
$${d\over dt}{\cal F}\bigl(m(\cdot,t)|m^c\bigr)=-{\cal I}
\bigl(m(\cdot,t)\bigr)$$ along the integral curves of (3.4).
\smallskip
\noindent {\it Proof}: We have, using (3.1) and (3.4)
$$\eqalign{
{\cal I}\bigl(m(\cdot,t)\bigr)&= -{d{\cal F}\over dt}=-\langle\partial_m{\cal F},{\partial m \over\partial
t}\rangle-\langle\overline\partial_m{\cal F},{\partial
\overline m\over\partial t}\rangle\cr &={1\over\beta}\re
\Sum_{x\in\Lambda^*} \bigl(-\beta J_\delta *\overline
m(x)+{\widehat I'(|m|)\over|m|} \overline m(x)\bigr)
\bigl(m(x)-f(\beta|J_\delta *m|){\beta J_\delta *m\over |\beta
J_\delta *m|} (x)\bigr)\cr} \leqno(3.5)$$
Let $m=\rho e^{i\theta}$, $\beta J_\delta *m=\rho' e^{i\theta'}$, ${\cal
I}\bigl(m(\cdot,t)\bigr)$ equals a sum of terms of the form
$$R={2\over\beta}\bigl(\rho'f(\rho')+\rho\widehat I'(\rho)
-\bigl(\rho\rho'+f(\rho')\widehat I'(\rho)\bigr)\cos
(\theta-\theta')\bigr)$$ then using
$\bigl(\rho-f(\rho')\bigr)\bigl(\widehat I'(\rho)-\rho'\bigr)\geq
0$ for any $\rho,\rho'$ since $\widehat I'$ is increasing, we
obtain the lower bound $R\geq{2\over\beta}\bigl(1-\cos
(\theta-\theta')\bigr) \bigl(\rho\rho'+f(\rho')\widehat
I'(\rho)\bigr)\geq 0$. And because $\rho\rho'+f(\rho')\widehat
I'(\rho)=0$ iff $\rho=0$ or $\rho'=0$, this estimate easily
implies the Proposition. $\clubsuit$
\medskip
From Proposition 3.1 and a compactness argument as in [Pr], follow
that in the closure of each orbit of equation (3.4) there is a
solution of (3.3), or equivalently, of Euler-Lagrange equation (3.2), i.e. a
critical point for ${\cal F}$. As suggested by numerical
simulations, this critical point is not unique, and depends on
initial conditions (except of course when $\beta\leq 2$.) We
expect however some uniqueness in the thermodynamical limit
$\Lambda^*\to\infty$, modulo the symmetry group.
\medskip
Now we give estimates on solutions of (3.4) or (3.3), borrowing
some ideas to [Pr]. Eq. (3.4) can be rewritten in the integrated
form~:
$$m(x,t)=e^{-t}m(x,0)+\int_0^t dt_1e^{t_1-t}f(\beta|J_\delta*m|)
{J_\delta*m\over|J_\delta*m|}(x,t_1)\leqno(3.6)$$ An effective
construction of the solution is given by the ``time-delayed''
approximations. It will also be used, discretizing time, in the
numerical simulations below. We define inductively $m_h(x,t)$,
$h>0$, on the intervals $[hk,h(k+1)[$, $k\in{\bf N}$, by
$m_h(x,t)=e^{-t}m_0(x)$ for $0\leq t<h$, and for $hk\leq
t<h(k+1)$, $k\geq 1$~:
$$m_h(x,t)=e^{kh-t}m_h(x,kh)+\int_{hk}^t dt_1e^{t_1-t}
f(\beta e^{-h}|J_\delta*m_h|)
{J_\delta*m_h\over|J_\delta*m_h|}(x,t_1-h)\leqno(3.7)$$ Using
Lipschitz properties of the coefficients, it is easy to see that,
as $h\to 0$, $m_h(x,t)$ tends to the solution $m(x,t)$ of (3.4)
uniformly for $x\in\Lambda^*$ and $t$ in compact sets of ${\bf
R}_+$. We prove estimates on $m(x,t)$ using sub- and
supersolutions of (3.4). We start with~:
\medskip
\noindent {\bf Lemma 3.2}: Assume $\beta> 2$, and let
$\lambda(t)$, $t>0$ be the solution of
$${d\lambda\over dt}(t)+\lambda(t)-f(\beta\lambda(t))=0, \ \lambda(0)=\lambda
\in[0,1[\leqno(3.8)$$
If $\lambda\geq m_\beta$, then
$\lambda(t)\leq\lambda$ for all $t>0$.
\smallskip
\noindent {\it Proof}: Write (3.8) in the integrated form as in
(3.6) and consider the approximating sequence $\lambda_h(t)$.
Since $\lambda_h(t)$ tends to $\lambda(t)$ uniformly on compact
sets of ${\bf R}_+$, it suffices to show the property stated in
the Lemma for $\lambda_h$, and $h>0$ small enough. For $0\leq
t<h$, $\lambda_h(t)=\lambda e^{-t}$, so the property holds, while
for $h\leq t< 2h$, performing the integration in (3.7), we get
$\lambda_h(t)=e^{-t}\lambda+\int_{h}^t dt_1e^{t_1-t} f(\beta
e^{-t_1}\lambda)$. Since $x>f(\beta x)$ iff $\beta x<\widehat I'(x)$ (whence iff $x>m_\beta$,~)
if $\lambda>m_\beta$, and $h>0$ small enough, then $f(\beta
e^{-t_1}\lambda)\leq e^{-t_1}\lambda$, and
$\lambda_h(t)\leq\lambda$. By induction, using also that $f$ is
increasing, but without changing $h>0$ anymore, it is easy to see
that this property carries over for all $t>0$. By a continuity argument, this holds true for all $\lambda\geq m_\beta$.
$\clubsuit$.
\medskip
Then we claim that the modulus of the magnetization doesn't
increase beyond $m_\beta$. More precisely we have~:
\medskip
\noindent {\bf Proposition 3.3}: Assume $\beta> 2$, and let
$m(x,t)$ be the solution of (3.4) such that $m_0(x)=m(x,0)$
satisfies $|m_0(x)|\leq\lambda<1$, for some $\lambda\geq m_\beta$,
and all $x\in{\bf Z}^2$ (so including the boundary condition on the exterior region~.) Then
$|m(x,t)|\leq\lambda$ for all $x\in{\bf Z}^2$, and all $t>0$.
\smallskip
\noindent {\it Proof}: Eq. (3.6) shows that
$$|m(x,t)|\leq e^{-t}|m_0(x)|+\int_0^t dt_1e^{t_1-t}
f(\beta|J_\delta*m|)(x,t_1)\leqno(3.9)$$ Now by the monotony
properties of the convolution and the function $f$, we have
$f(\beta|J_\delta*m|)(x,t_1)\leq f(\beta J_\delta*|m|)(x,t_1)$, so
the solution $\lambda(t)$ of (3.8) with $\lambda(0)=\lambda$ is a
supersolution for (3.9), and Lemma 3.2 easily implies the
Proposition. $\clubsuit$
\medskip
We now look for lower bounds on $m(x,t)$. Since there are in
general vortices, one cannot expect a global, positive lower bound
on $|m(x,t)|$, unless there is no vorticity on initial and
boundary values. On the other hand, we know (at least for a 1-d
lattice, see [BuPi],~) that the Gibbs measure of the
configurations at equilibrium $m_\delta\in\widetilde{\cal X}
(\Lambda^*|\Lambda^{*c})$ with $|m_\delta(x)|$ arbitrarily close
to $m_\beta$, has to be large. We have~:
\medskip
\noindent {\bf Proposition 3.4}: Assume $\beta> 2$, and let
$m(x,t)$ be the solution of (3.4) such that $m_0(x)=m(x,0)$ as in
Proposition 3.3 satisfies $\re (\nu m_0(x))>\mu$, for some fixed
$\nu\in{\bf S}^1\approx\{z\in{\bf C},|z|=1\}$ and $\mu>0$ and all $x\in{\bf Z}^2 $. Assume
furthermore that $\mu$ satisfies $(\mu^2+\lambda^2)^{1/2}<\beta
f(\beta\lambda)$, where $\lambda$ is as in Proposition 3.3. Then
$\re (\nu m(x,t))\geq\mu$ for all $x\in\Lambda^*$, and all $t>0$.
\smallskip
\noindent {\it Proof}: As in the proof of Proposition 3.3 we make
use of a comparison function. So let $\mu(t)$ verify the
differential equation
$${d\mu\over dt}(t)+\mu(t)-\beta f(\beta\lambda)
{\mu(t)\over(\mu(t)^2+\lambda^2)^{1/2}}=0, \ \mu(0)>\mu
\leqno(3.10)$$
Write (3.10) in the integrated form as in (3.6) and
consider the approximating sequence $\mu_h(t)$ as in (3.7). We
shall show that $\mu_h(t)\geq\mu$ for all $t>0$ provided
$\mu(0)>\mu$ verifies the inequality given in the Proposition.
Namely, this holds for $0\leq t<h$, because then
$\mu_h(t)=\mu(0)e^{-t}\geq\mu$ for $h>0$ small enough, while  for
$h\leq t< 2h$, performing the integration as in (3.7), we get
$\mu_h(t)=e^{h-t}\mu(0)+\beta f(\beta\lambda)\int_h^t dt_1
e^{t_1-t}{\mu e^{-t_1}((\mu e^{-t_1})^2+\lambda^2)^{-1/2}}$. By
hypothesis, ${\mu e^{-t_1}\beta f(\beta\lambda)((\mu
e^{-t_1})^2+\lambda^2)^{-1/2}}\geq \mu$ for $h$ small enough. So
again $\mu_h(t) \geq \mu$. By induction, using that the function
$\rho\mapsto\rho(\rho^2+\lambda^2)^{-1/2}$ is increasing on ${\bf
R}_+$, it is easy to see that $\mu_h(t)\geq \mu$ holds for all
$t>0$. Because the coefficients of (3.10) are uniformly Lipschitz,
$\mu_h(t)$ tends to $\mu(t)$ uniformly on compact sets in ${\bf
R}_+$, and this property holds again for $\mu(t)$.

Now we turn to the equation for $m(x,t)$. Possibly after rotating
the coordinates, we may assume $\nu=1$, i.e. $\re m_0(x)\geq\mu$
and all $x\in\Lambda^*$ (again, we have included the boundary
condition in the initial configuration.~) Write $m(x,t)=u(x,t)+
iv(x,t)$, $u,v$ real and take real part of (3.4). The integrating
form of the resulting equation writes~:
$$u(x,t)=e^{-t}u(x,0)+\int_0^t dt_1e^{t_1-t}f(\beta|J_\delta*m|)
{\beta J_\delta*u\over\beta |J_\delta*m|}(x,t_1)\leqno(3.11)$$ As
$\rho'\mapsto {f(\rho')\over\rho'}$ is decreasing on ${\bf R}_+$,
and by Proposition 3.2, $|J_\delta*m|\leq\bigl(|J_\delta*u|^2
+\lambda^2\bigr)^{1/2}$, we have
$${f(\beta|J_\delta*m|)\over\beta |J_\delta*m|}
\geq{f\bigl(\beta\bigl(|J_\delta*u|^2
+\lambda^2\bigr)^{1/2}\bigr)\over\beta\bigl(|J_\delta*u|^2
+\lambda^2\bigr)^{1/2}}\geq {f(\beta\lambda)\over
\beta\bigl(|J_\delta*u|^2 +\lambda^2\bigr)^{1/2}}$$ the last
inequality because $f$ is increasing. Since $u(x,0)\geq\mu$, by
continuity we have $u(x,t)> 0$ at least for small $t>0$, and
(3.11) gives
$$u(x,t)\geq e^{-t}u(x,0)+\beta f(\beta\lambda)
\int_0^t dt_1e^{t_1-t}(J_\delta*u)
\bigl((J_\delta*u)^2+\lambda^2\bigr)^{-1/2}(x,t_1)\leqno(3.12)$$
Now, using the monotony of the convolution, and again the fact
that the function $\rho\mapsto\rho(\rho^2+\lambda^2)^{-1/2}$ is
increasing on ${\bf R}_+$, we can easily show that the solution
$\mu(t)$ of (3.10) with $\mu(0)=\mu$ is actually a subsolution for
(3.12), for all $t>0$~; the properties proved already for $\mu(t)$
then imply the Proposition. $\clubsuit$
\medskip
Of course, by continuity, Propositions 3.3 and 3.4 imply the
corresponding estimates for the solutions of (3.3), or equivalently for the solutions of
Euler-Lagrange equation (3.2). Our last result states that if $\beta\leq
2$, then $m(x,t)$ tends to 0 $t\to\infty$, which is consistent
with the absence of phase transition (or spontaneous
magnetization) at high temperature.
\medskip
\noindent {\bf Proposition 3.5}: Assume $\beta\leq 2$, and let
$m(x,t)$ be the solution of (3.4). Then $m(x,t)\to 0$ on
$\Lambda^*$ as $t\to+\infty$.
\smallskip
\noindent {\it Proof}: Using that $f(\rho')\leq{1\over 2}\rho'$,
all $\rho'>0$, (3.9) shows that
$$|m(x,t)|\leq e^{-t}|m_0(x)|+{\beta\over 2}\int_0^t dt_1e^{t_1-t}
J_\delta*|m|(x,t_1)$$ So by taking convolution
$$J_\delta*|m|(x,t_1)\leq e^{-t_1}J_\delta*|m_0|(x)+{\beta\over 2}
\int_0^{t_2} dt_2e^{t_2-t_1} J_\delta^{*2}*|m|(x,t_2)$$ and
integrating the resulting inequality~:
$$|m(x,t)|\leq e^{-t}\bigl[|m_0(x)|+{\beta t\over 2}J_\delta*|m_0|(x)
+\bigl({\beta \over 2}\bigr)^2T^{(2)}
\bigl(e^{(\cdot)}J_\delta^{*2}*|m|(x,\cdot)\bigr)(t)\bigr]$$ where
$T^{(k)}u(t)=\int_0^tdt_1\int_0^{t_1}dt_2\cdots
\int_0^{t_{k-1}}dt_ku(t_k)$ denotes the $k$-fold integral of $u$,
and $J_\delta^{*k}$ the $k$-fold convolution product of $J_\delta$
with itself. By induction, we get~:
$$\eqalign{
|m(x,t)|&\leq e^{-t}\bigl[ |m_0(x)|+{\beta \over 2}t
J_\delta*|m_0|(x) +\cdots+\bigl({\beta \over 2}\bigr)^k {t^k\over
k!}J_\delta^{*k}*|m_0|(x)\cr &+T^{(k+1)}
\bigl(e^{(\cdot)}J_\delta^{*(k+1)}*|m|(x,\cdot)\bigr)(t)\bigr]\cr
}$$ The series is uniformly convergent for $t$ in compact sets so
we can write
$$|m(x,t)|\leq e^{-t}\Sum_{k=0}^{+\infty}
\bigl({\beta \over 2}\bigr)^k {t^k\over
k!}J_\delta^{*k}*|m_0|(x)$$ When $\beta<2$, using
$J_\delta^{*k}*|m|(x,0)\leq |m_0(x)|\leq 1$, it follows that
$m(x,t)\to 0$ for all $x\in\Lambda^*$ as $t\to\infty$. This holds
again for $\beta=2$ since we may assume that $m_0$ has compact
support, and we know (see [H\"o,Lemma 1.3.6]) that
$J_\delta^{*k}\to 0$ uniformly on ${\bf R}^2$ (or on ${\bf Z}^2$
in the discrete case,~) as $k\to\infty$. $\clubsuit$

\medskip
\noindent {\bf 4. Vortices}.
\medskip
We consider here the problem of finding numerically the critical
points of Euler-Lagrange equation (3.3) by solving (3.4) subject
to a boundary condition on $\Lambda^{*c}$ presenting vorticity.
\medskip
\noindent{\bf a) Generalities}.
\smallskip
First we recall some facts about the degree of a map. Let $m:{\bf
R}^2\to {\bf C}$ be a differentiable function, considered as a
vector field on ${\bf R}^2$,  and subject to the condition
$|m(x)|\to \ell >0$ as $|x|\to\infty$ uniformly in $\widehat
x=x/|x|$. Then the integer
$$\deg _{R} m={1\over 2\pi}\int_{|x|=R}d(\arg m)={1\over 2i\pi}\int_{|x|=R}{dm\over m}\leqno(4.1)$$
is independent of $R$ when $R>0$ is large enough, is called the
(topological) degree of $m$ at infinity, and denoted by
$\deg_\infty m$.

We define in the same way the local degree (or topological defect)
$\deg _{x_0} m$ of $m$ near $x_0$, provided $m(x)\neq 0$, $x\neq
x_0$, by integrating on a small loop around $x_0$. The local
degree takes values $d_j\in{\bf Z}$. When $m$ has finitely many
zeros $x_j$ inside the disc of radius $R$, its total degree (or
vorticity) is defined again as the sum of all local degrees near
the $x_j$'s. In many boundary value problems (or generalized
boundary value problems, in the sense that the boundary is at
infinity,~) such as Ginzburg-Landau equations, total vorticity is
conserved, i.e. $\deg _\infty m=\Sum_j \deg _{x_j}m$. Generically
$d_j=\pm 1$ (``simple poles''.) Our aim is to check this
conservation principle in the present situation.

We can define analogously the degree of a discrete map, which
makes sense at least in the thermodynamical limit. If
$m(x)=\rho(x) e^{i\theta(x)}$, the degree of $m$ at infinity is
the degree restricted to the lattice $\Lambda^{*c}$, e.g. by
$$d=\deg _{\Lambda^{*c}}m={1\over 2\pi}\Sum_j (\theta_{j+1}-\theta_j)\leqno(4.2)$$
along some closed loop $\Gamma_\iota\subset\Lambda^{*c}$
encircling $\Lambda^*$, the sites along $\Gamma_\iota$ being
labelled by $j$, assuming that this integer takes the same value
on each $\Gamma_\iota$.

The local degree near $x_0$, where $m(x_0)=0$, is identified again
by computing the angle circulation on a loop encircling $x_0$.
Local degrees are also expected to take, generically, values $\pm
1$.

We chose our parameters as follows. We start with prescribing the
degree of the spin variable $\sigma$ on $\Lambda^c$, and take on
$\Gamma_\iota$, the $\iota$:th loop away from $\Lambda$,
containing $N_\iota$ sites, ($N_\iota=4\iota+P$, where $P$ is the
perimeter of $\Lambda$, we take enough $\iota$'s to cover the
range of interaction,~) with a uniform distribution:
$$\sigma_j=\exp i(2\pi d j/N_\iota+\phi_0),\ 1\leq j\leq
N_\iota\leqno(4.3)$$ here $\phi_0$ is a constant (e.g. $\phi_0=1$)
that ``breaks'' the symmetry of the rectangle $\Lambda$. We shall
also randomize these boundary conditions.

To this spin distribution on $\Lambda^c$, we apply the block spin
transformation (1.3), so to have a distribution of magnetization
on $\Lambda^{*c}$, then we prescribe  initial conditions inside
$\Lambda^*$. The simplest way is to take zero initial values,
which  gives a particular symmetry to the solution. Otherwise, we
can choose them as random numbers, either small, or with absolute
value less than $m_\beta$. All these cases will be discussed.

We usually fix the inverse temperature $\beta=5$, so
$m_\beta=0.72$~; the results do not depend on $\beta$ in an
essential way, we just observe that magnetization tends to $0$ as
$\beta \to 2^+$. The diameter $L$ of the lattice $\Lambda$ ranges
from $2^6$ to $2^{10}$, the size $\delta/\gamma$ of the diameter
of the block-spin $\Delta(x)$ is set to 4 (most of the time) so
the diameter $L^*$ of the lattice $\Lambda^*$ ranges from $2^4$ to
$2^{8}$. The lattice is either a square, or a rectangle.

The size $1/\delta$ of the length of interaction in $\Lambda^*$
ranges from 2 to 32, thus the corresponding interaction in
$\Lambda$ has length $1/\gamma = 4/\delta$ between 8 to 128.

Equation (3.4) is solved by ``time-delayed'' approximations as in
(3.7), implemented by the second order trapezoidal method to
compute the integrals.

These experiments lead to the following observations, vortices
display in a different way, according to the initial configuration
on $\Lambda^*$, but always obey the conservation of total
vorticity.
\medskip
\noindent {\bf b) Some typical configurations}.
\smallskip
We consider here the case of a uniform distribution of spins on
the boundary.

The particular case of zero initial values and a square lattice,
gives raise to interesting symmetries (or degeneracies) in the
picture~: namely, vortices tend to occupy most of $\Lambda^*$ so
to cope with the symmetry of the square. So for $d=1$ there is a
single vortex in the center, for $d=2$ (cf Fig 1.a) a vortex of
multiplicity 2, (unless the degeneracy is lifted and turns into 2
nearby vortices,~) for $d=4-1$, (cf Fig 2.a) one vortex of degree
-1 surrounded by 4 vortices of degree +1 near the corners, for
$d=4$, 4 vortices of degree +1 near the corners, for $d=4+1$, same
configuration as for $d=3$, for $d=4+2$ the picture looks alike,
with a double vortex at the center, for $d=2\times 4-1$, 4 new
vortices appear near the center (cf Fig 3.a), etc\dots So the
configuration depends essentially of the residue of $d$ modulo 4~:
new vortices show up from the middle towards the corners along the
diagonals of $\Lambda^*$.

\centerline{\includegraphics[width=8cm,height=6cm]{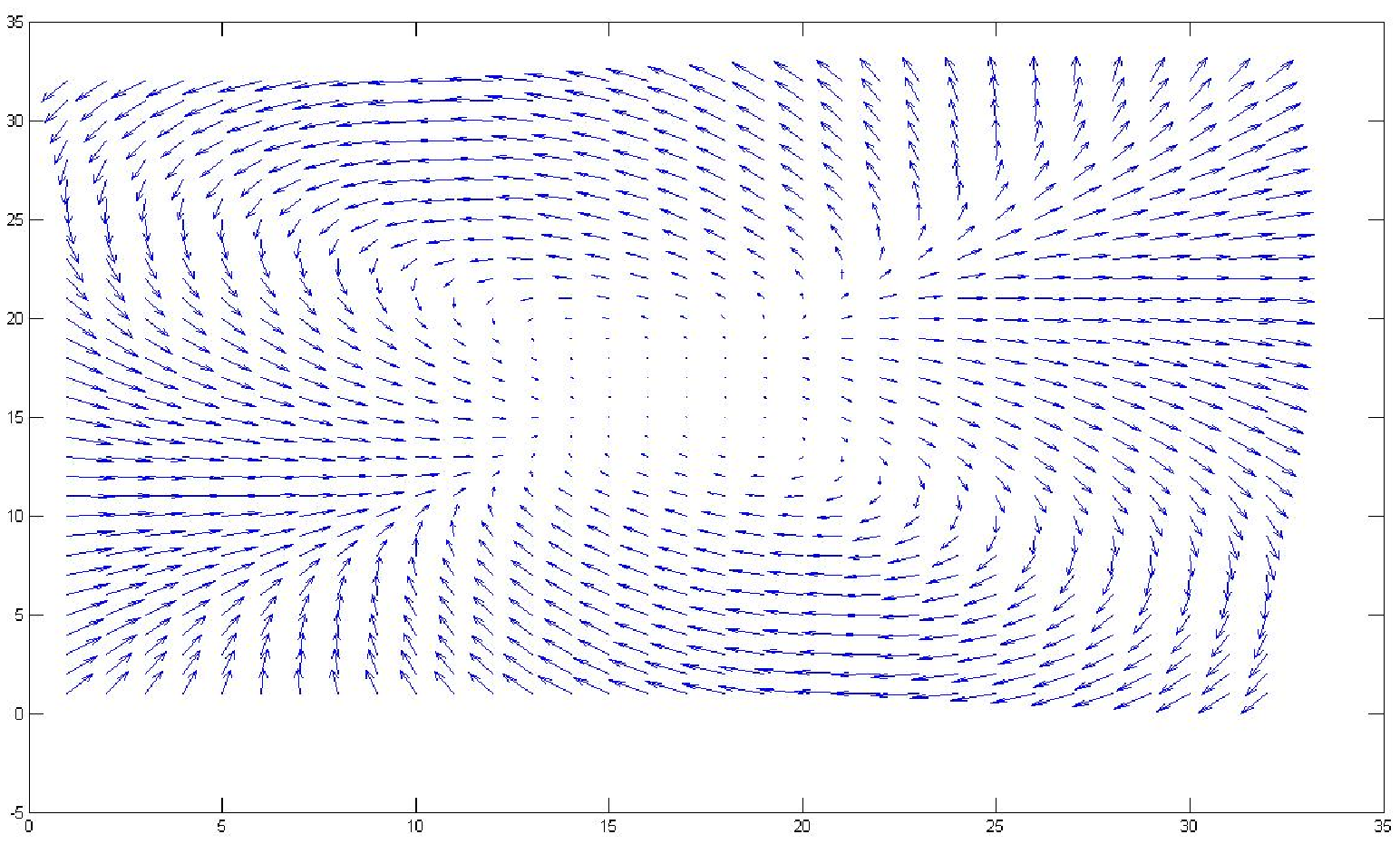}
\includegraphics[width=8cm,height=6cm]{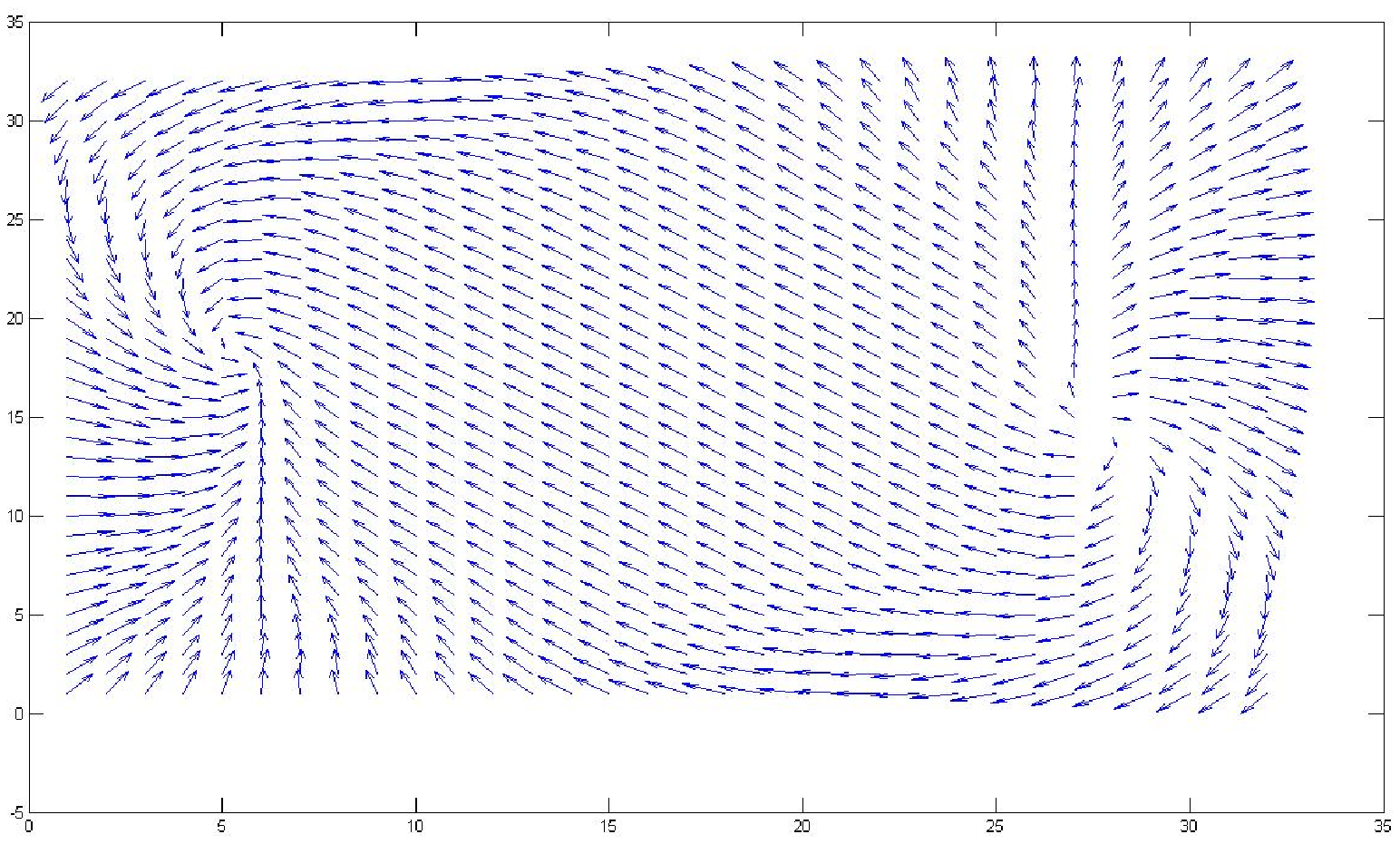}}
\leftline{\sevenrm {\sevenbf Fig 1.a}: $L^*=128, d=2$, zero
initial condition \ \ \ \ \ \ \ \ \ \ \ \ \ \  {\sevenbf Fig 1.b}: $L^*=128, d=2$, random initial condition}

\centerline{\includegraphics[width=8cm,height=6cm]{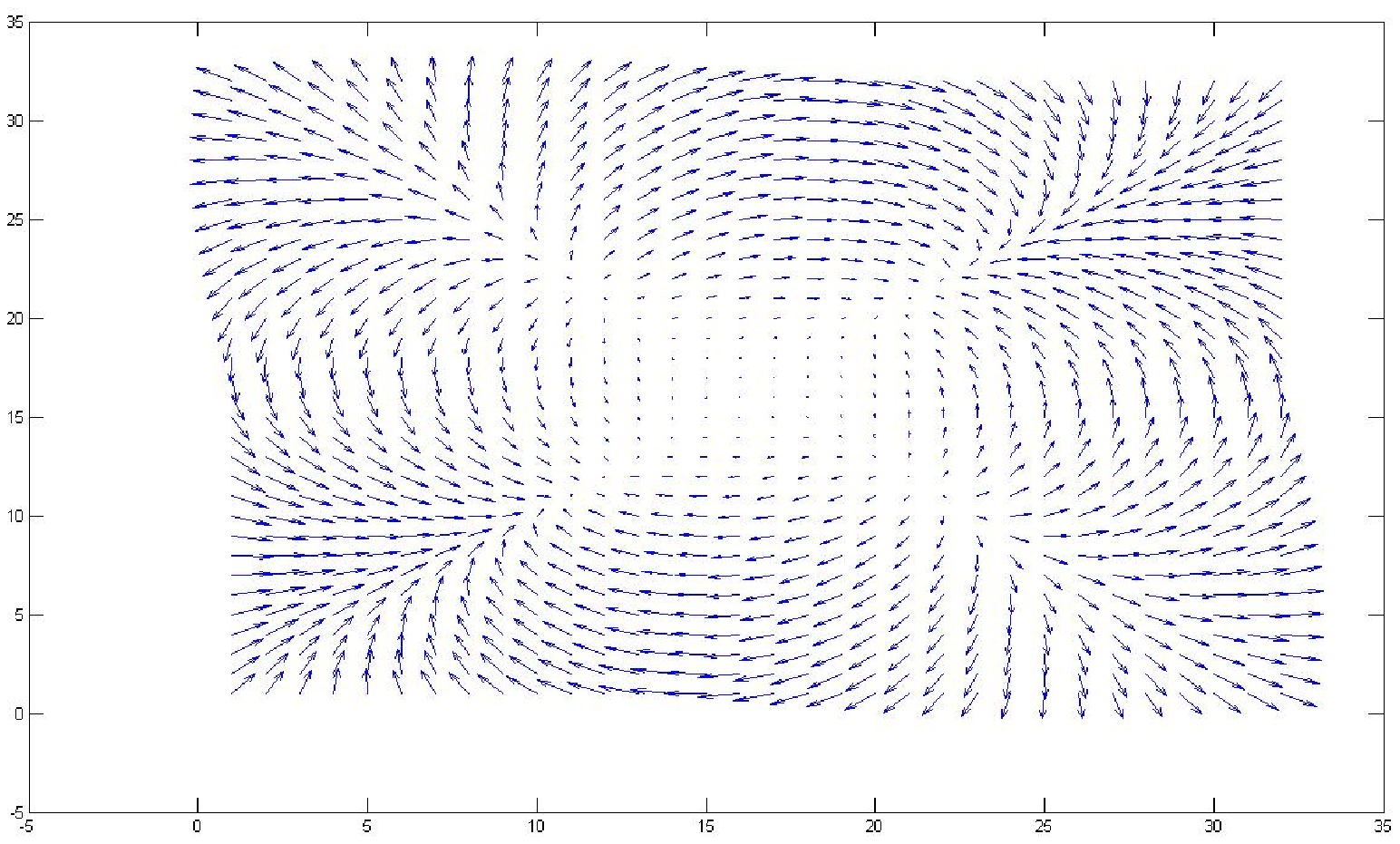}
\includegraphics[width=8cm,height=6cm]{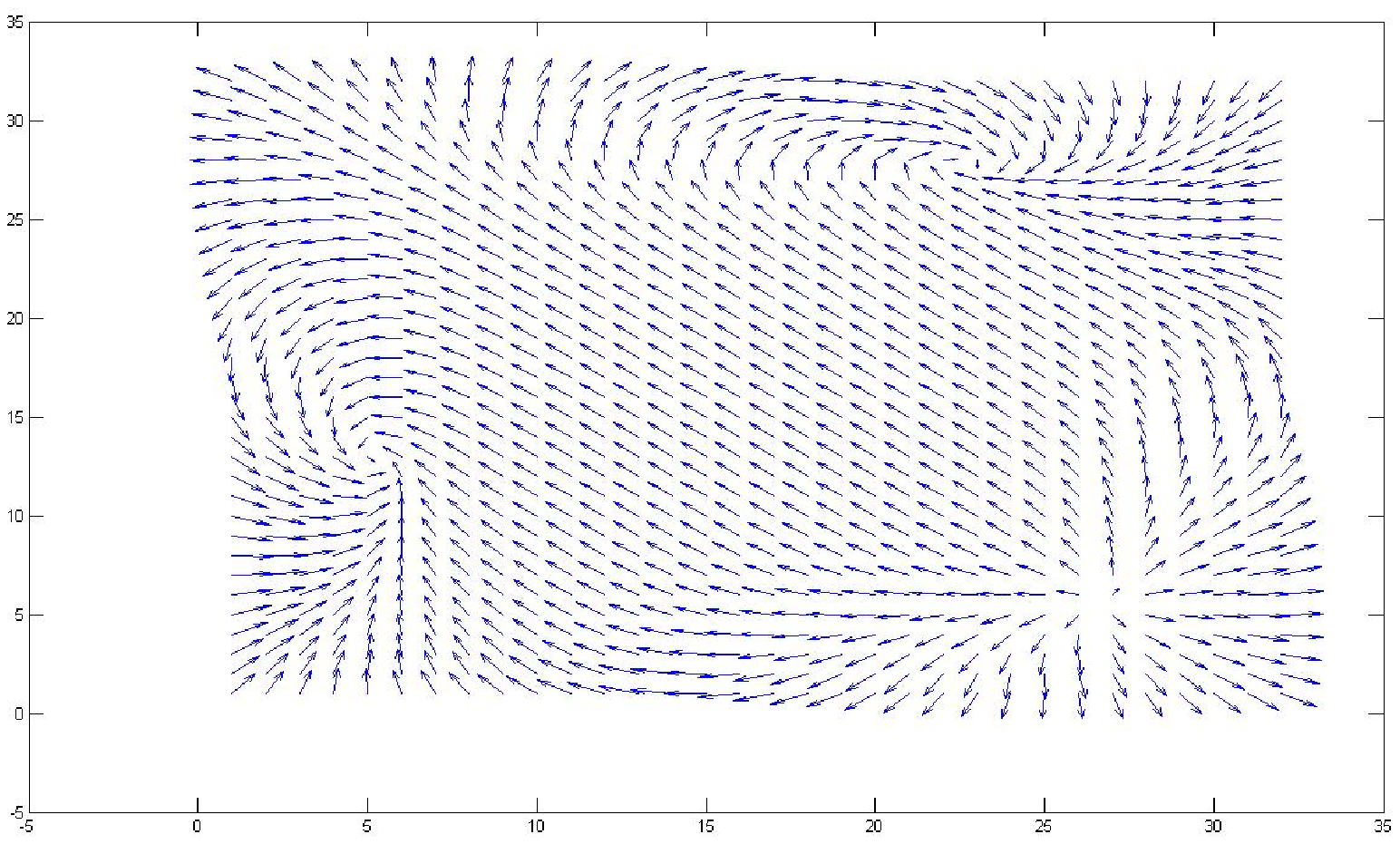}}
\leftline{\sevenrm {\sevenbf Fig 2.a}: $L^*=128, d=3$, zero
initial condition \ \ \ \ \ \ \ \ \ \ \ \ \ \ {\sevenbf Fig 2.b}:
$L^*=128, d=3$, random initial condition}

Next we consider the case of a square lattice, but with random
initial conditions, that is, we pick initial magnetizations with
random direction and random length, provided the length is much
smaller than $m_\beta$, typically $|m_0(x)|\leq 0.05$. Then
vortices are simple (i.e. have local degree $\pm 1$, total
vorticity is of course conserved,~) and tend to display at the
periphery of $\Lambda^*$, in a pretty regular way, leaving some
large ordered domain near the center.

Thus, these  configurations maximize the area of the lattice where
the magnetizations are aligned , with an absolute value close to
$m_\beta$, (in accordance with the fact that energy $H_\gamma$
decreases as the spins align.~) Their direction, in general,
points out along one of the diagonals of $\Lambda^*$. This is
illustrated in Fig.1,2,3.b above, for a vorticity $d=2,3,7$
respectively. In  particular, Fig.2 shows the topological
bifurcation from d=4-1 to d=3. These simulations also suggest that
the equilibrium configurations depend on the initial conditions,
but exceptional configurations due to symmetry, for zero initial
conditions, are essentially removed as soon as a small disorder is
introduced.

\centerline{\includegraphics[width=8cm,height=6cm]{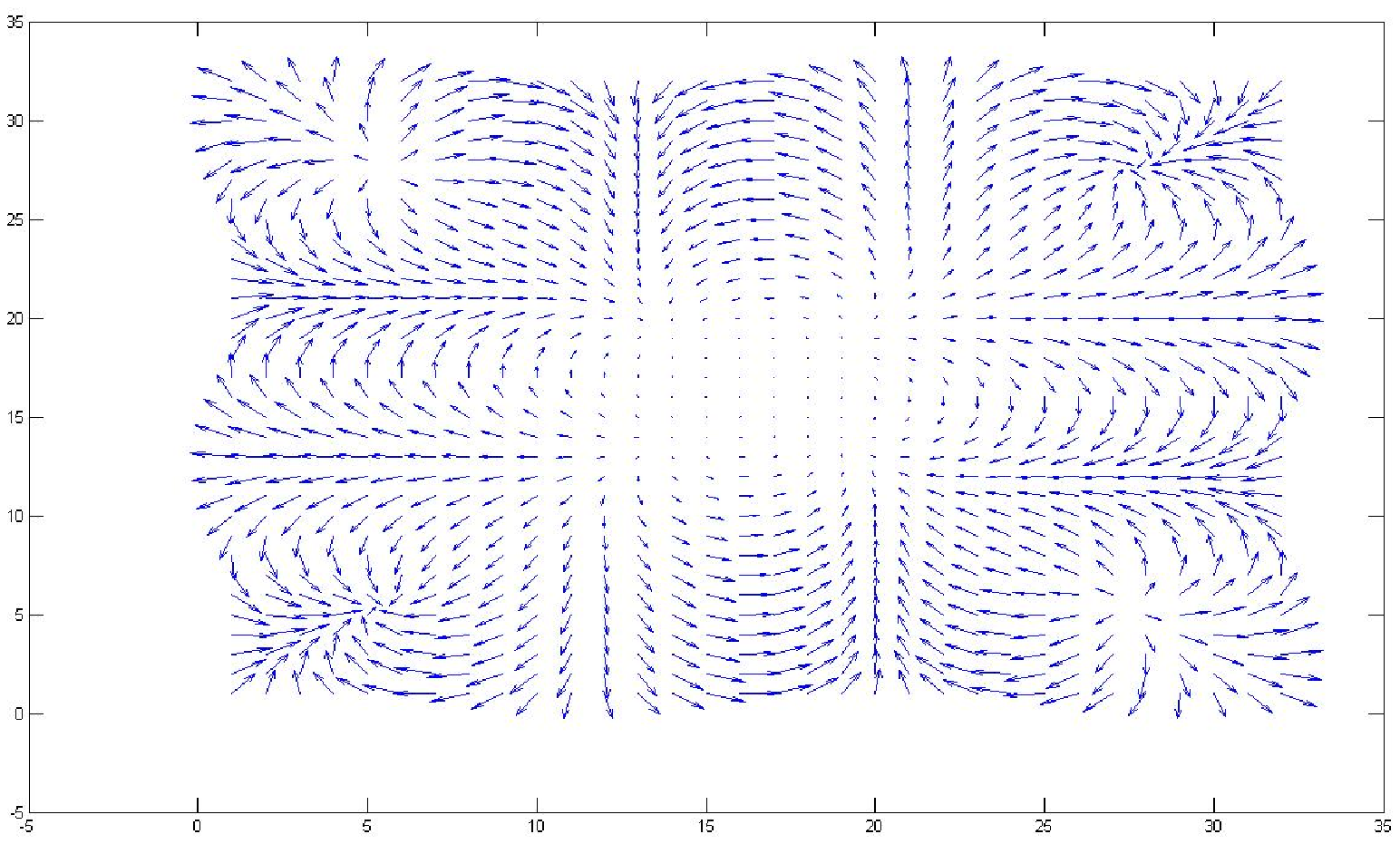}
\includegraphics[width=8cm,height=6cm]{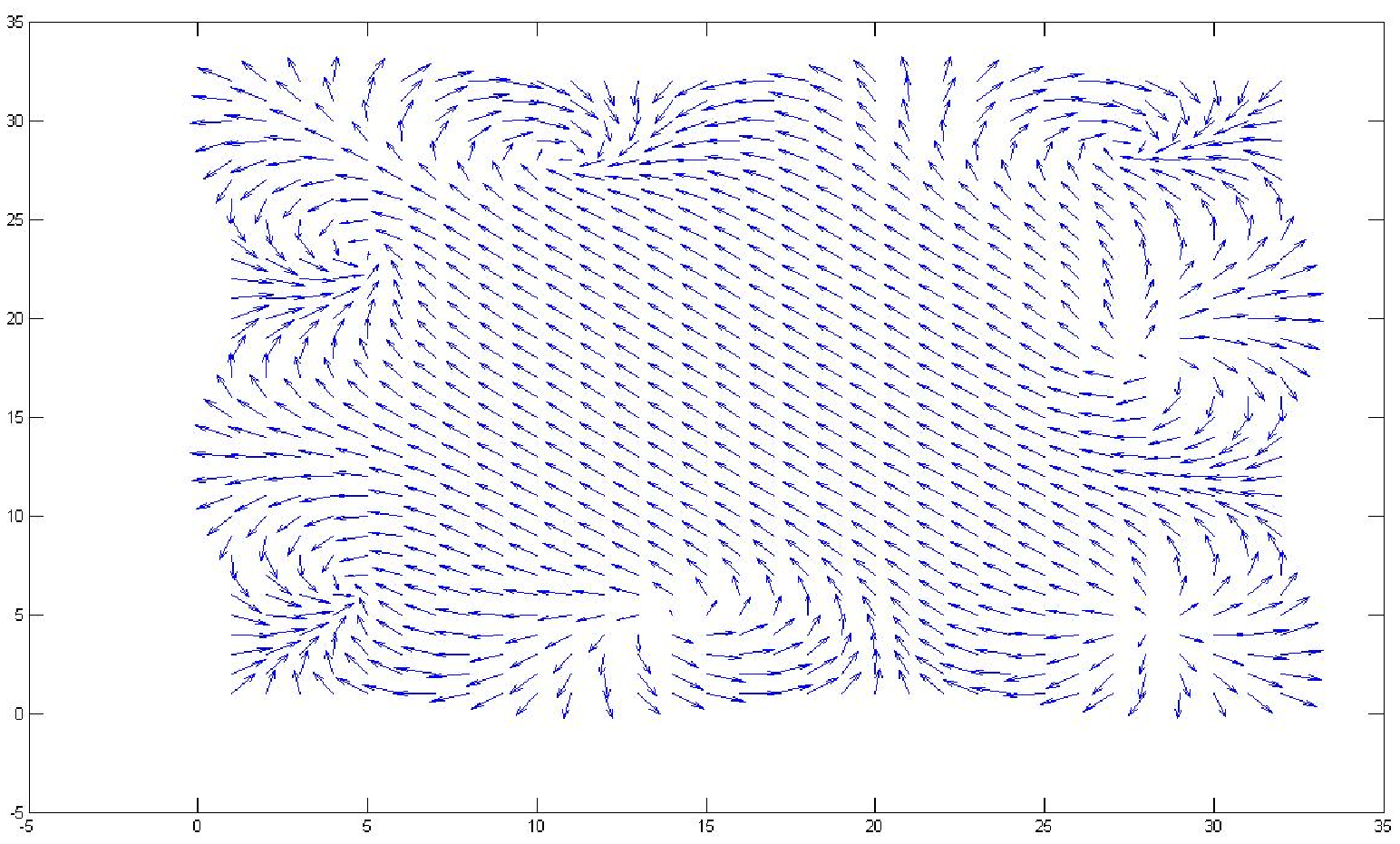}}
\leftline{\sevenrm {\sevenbf Fig 3.a}: $L^*=128, d=7$, zero
initial condition \ \ \ \ \ \ \ \ \ \ \ \ \ \ {\sevenbf Fig 3.b}:
$L^*=128, d=7$, random initial condition}

Now we vary the shape of the lattice, changing the square into a
rectangle, keeping in mind that thermodynamic limit, most of the
time, should be taken in the sense of Fisher, i.e. the length of
the rectangle $\Lambda^*$ doesn't exceed a constant times
$|\Lambda^*|^{1/2}$. As expected, vortices tend to align along the
largest dimension, but again, limiting configurations depend on
whether the initial condition inside $\Lambda^*$ is set to zero or
not.

\centerline{\includegraphics[width=15cm,height=8cm]{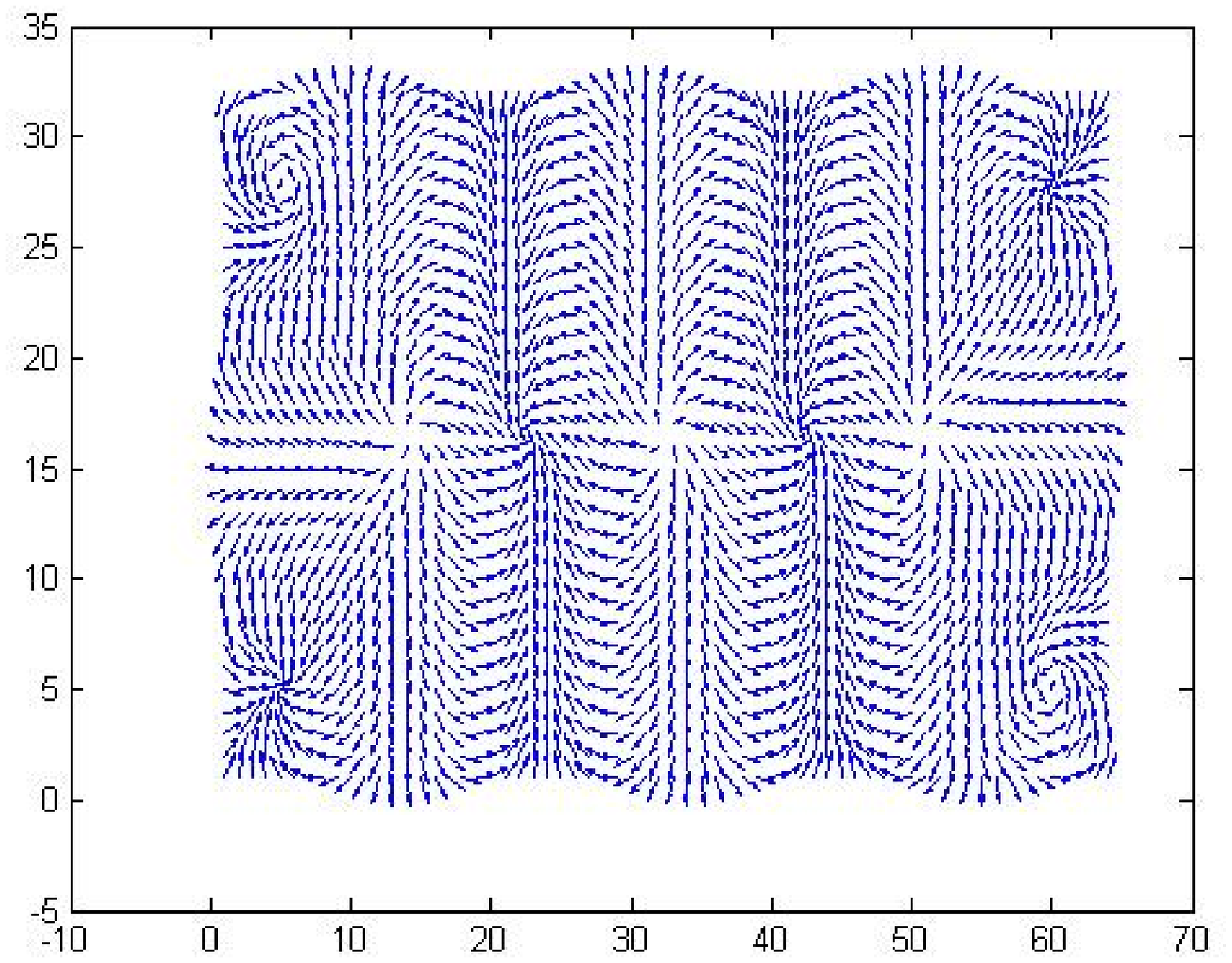}}
\centerline{\sevenrm {\sevenbf Fig 4.a}: $L^*=256, \ell^*=128,
d=9$, zero initial condition }

\centerline{\includegraphics[width=15cm,height=8cm]{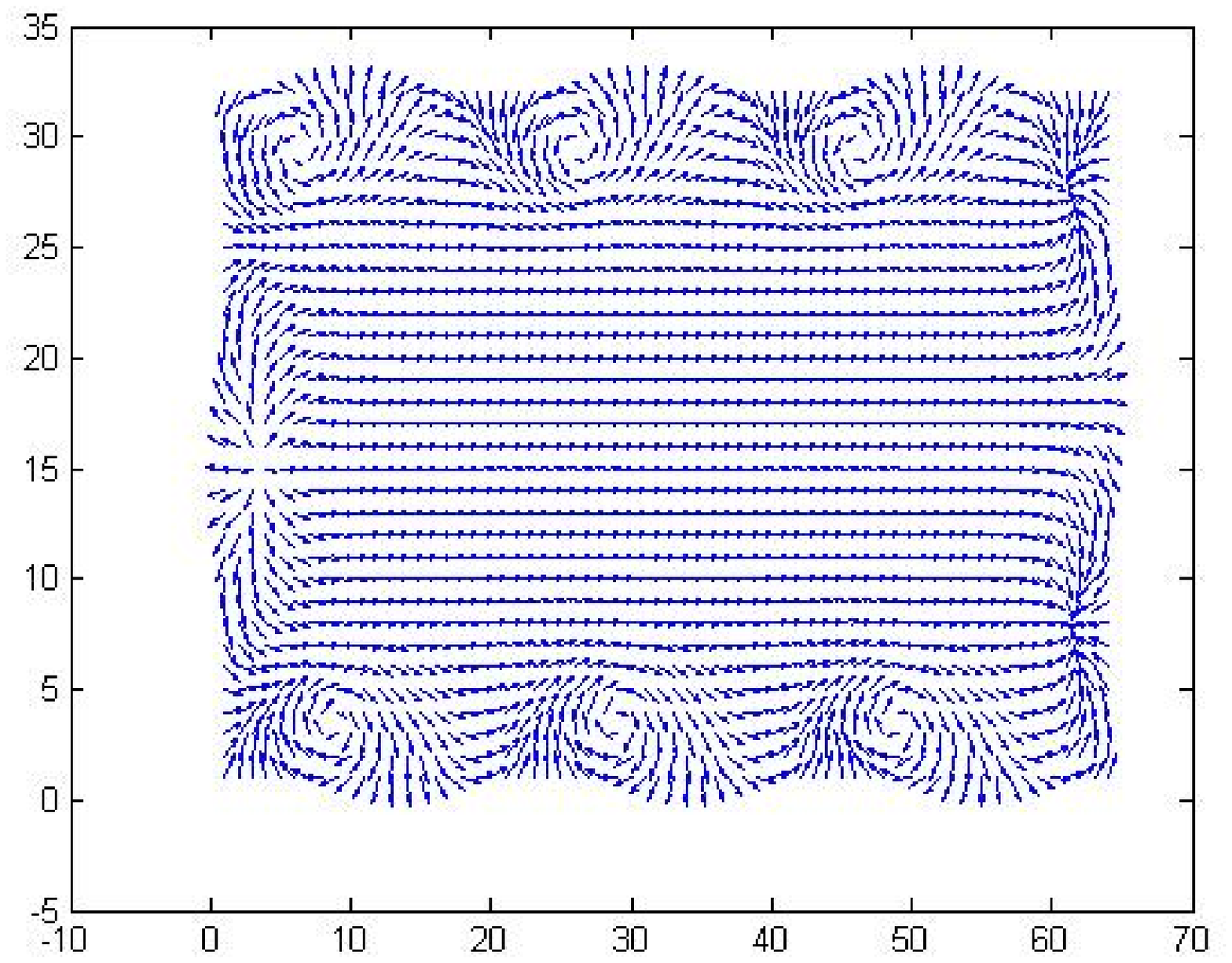}}
\centerline{\sevenrm{\sevenbf Fig 4.b}: $L^*=256,\ell^*=128, d=9$,
random initial condition}

\centerline{\includegraphics[width=15cm,height=8cm]{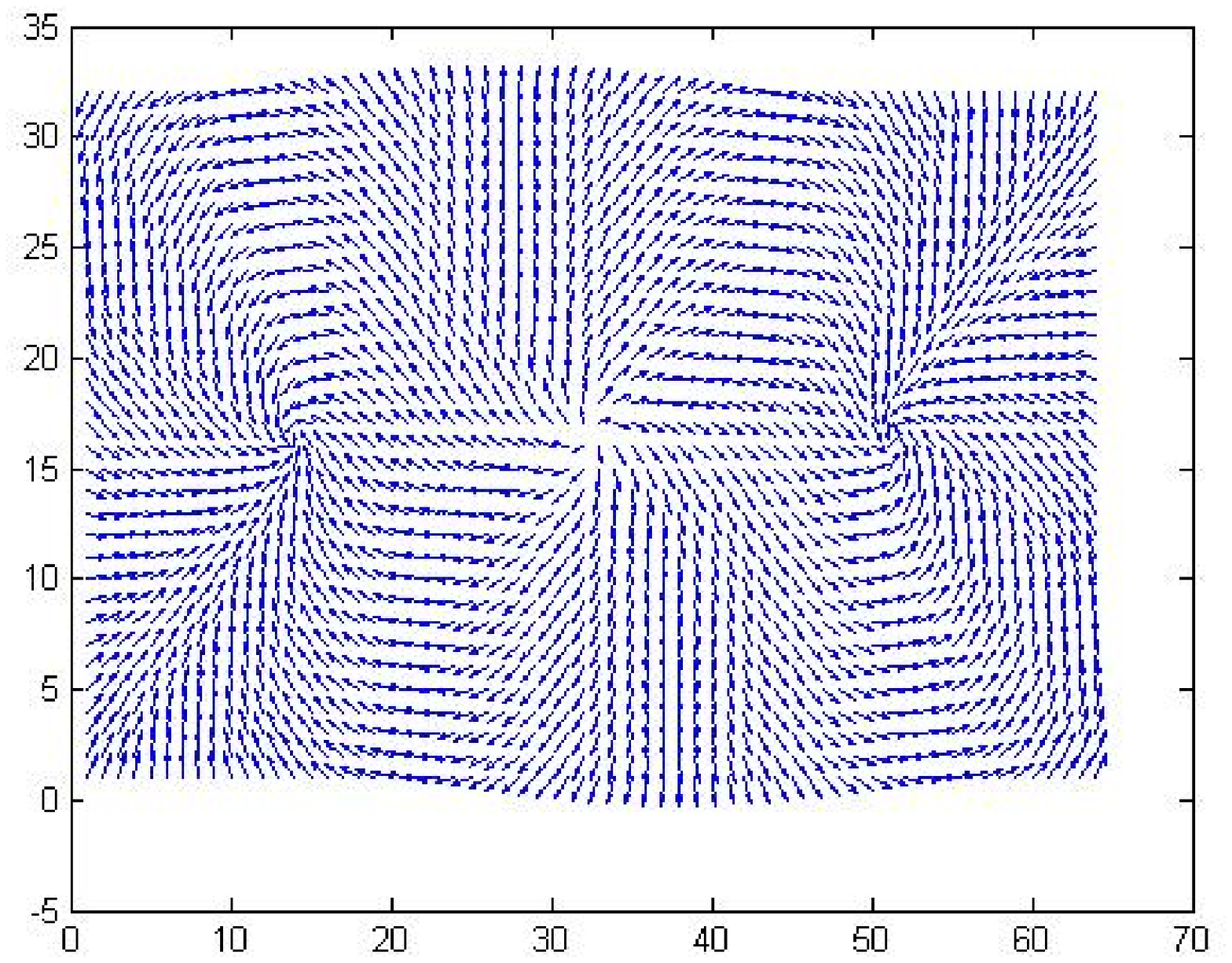}}
\centerline{\sevenrm {\sevenbf Fig 4.c}: $L^*=256,\ell^*=128,
d=3$, zero initial condition }

Thus, for zero initial condition, vortices display along the
largest median of $\Lambda^*$, with possible extra vortices near
the corners (inheriting the features of the square lattice.~)
Namely, they tend to repel each other so the energy cost in
clustering is minimized by occupying the corners. Typically, such
configurations occur if $d\geq 4$  and the length of $\Lambda^*$
is only twice its width. But for sufficiently long lattices, or
small degree, they just stand the median line. See Fig.4.a and
4.c.

For small random initial conditions as above (Fig.4.b), we recover
the general picture of square lattices, i.e. vortices set along
the boarder of $\Lambda^*$, leaving a large space in the middle
with parallel magnetizations. In any case, degeneracies are
lifted, and all vortices have degree +1.
\medskip
\noindent{\bf c) The simulated annealing}.
\smallskip
If we increase the initial conditions, still keeping $|m_0(x)|\leq
m_\beta$, we obtain similar pictures, but with a non uniform
distribution of defects~: conservation of total degree holds, but
at the same time, many vortices spread over the lattice, and the
corresponding long range order region shrinks correspondingly.
This suggest that the gradient-flow dynamics converges only to a
local minimum of the free energy.

For reaching lower energies, we let the system explore other
regions of the configuration space. This can be achieved through
simulated annealing, see e.g. [KiGeVec]. Replace the dynamics
(3.4) by
$${dm\over dt}=-m+f(\beta(t)|J_\delta *m|){J_\delta *m\over |J_\delta *m|}
\ \hbox{in} \ \Lambda^* \leqno(4.4)$$ where $\beta(t)$ depends
continuously on $t\in[0,t_1]$, starting with
$\beta_0<\beta_1=\beta$, with negative slope at $t=0$, so that the
system is heated initially up to a peak $\beta_2^{-1}\approx 1/2$
(the critical temperature) around $t=t_2$, and then gradually
cooled down to $\beta$ at $t=t_1$. Function $\beta(t)$ is
oscillating between successive warm and cool periods, so to
''shake'' sufficiently the system. Then we keep the temperature
constant till we reach equilibrium.

It is not difficult to optimize, empirically, the annealing
function $\beta(t)$, and our choice was the following~:

\centerline{\includegraphics[width=10cm,height=5cm]{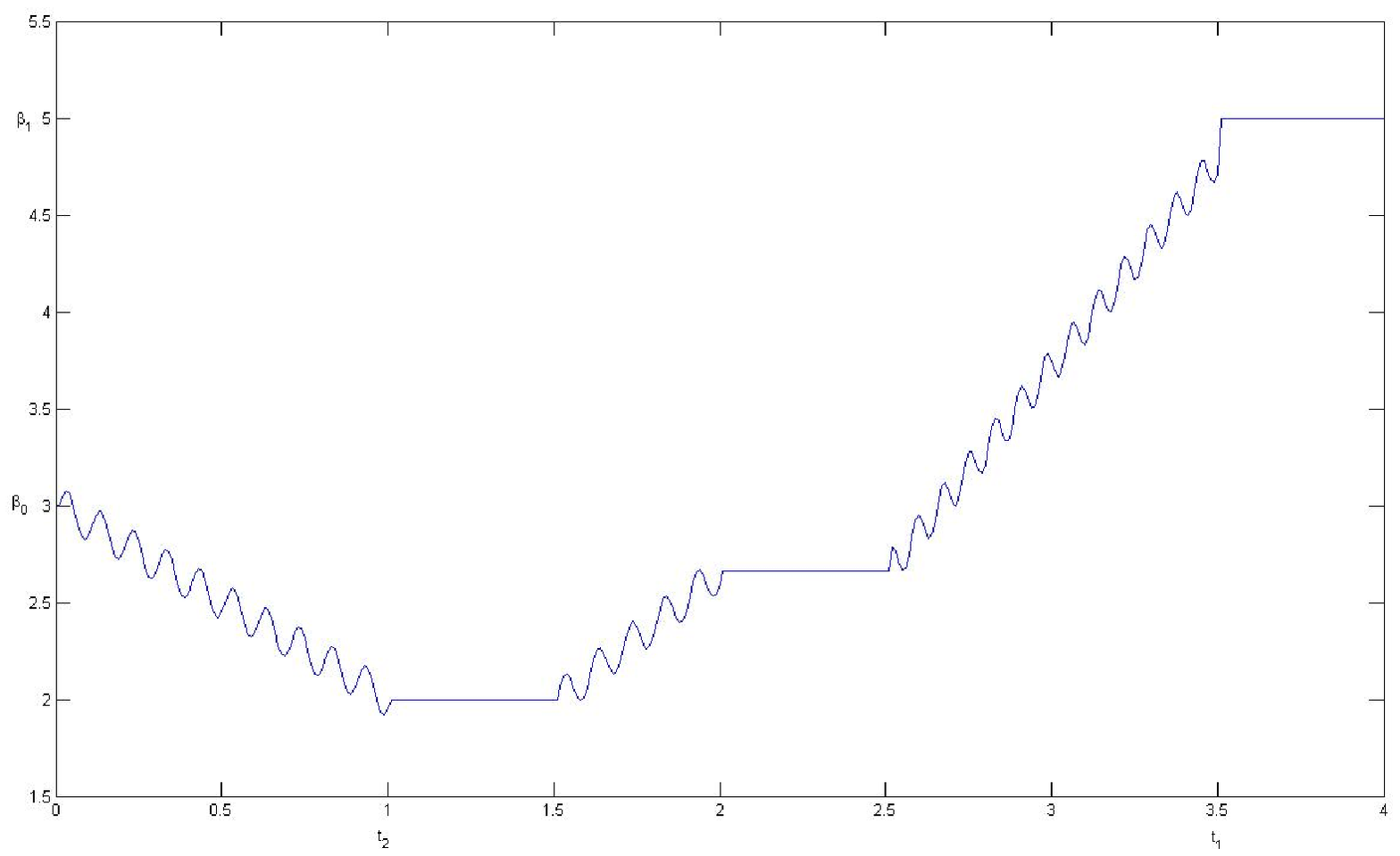}}
\centerline{\sevenrm {\sevenbf Fig 5}: the annealing function
$\beta(t)$ }

We applied this method first to the case of a square lattice, when
the equilibrium configuration corresponding to some total degree
$d_0$ is used as an initial condition for a dynamics with degree
$d_1$. We fix $\beta_1=5$, $L^*=128$.

Consider first the case $d_1=3$, the equilibrium configuration,
with 0 initial condition, is given in Fig. 2a, and the
corresponding free energy is $E=99$. We use simulated annealing to
compute the equilibrium, starting from $d_0=4,-3,5$, and find
respectively $E=23,53,51$, see Fig. 6. So the energies obtained
this way are less than with zero initial conditions, though the
initial magnetizations are rather large. At the same time,
symmetries get lost. Thus the cost for the 3 vortices to be
aligned along one of the diagonals of the square as in Fig 6.b is
less than to form a domino near the center as in Fig. 2a.

\centerline{\includegraphics[width=8cm,height=6cm]{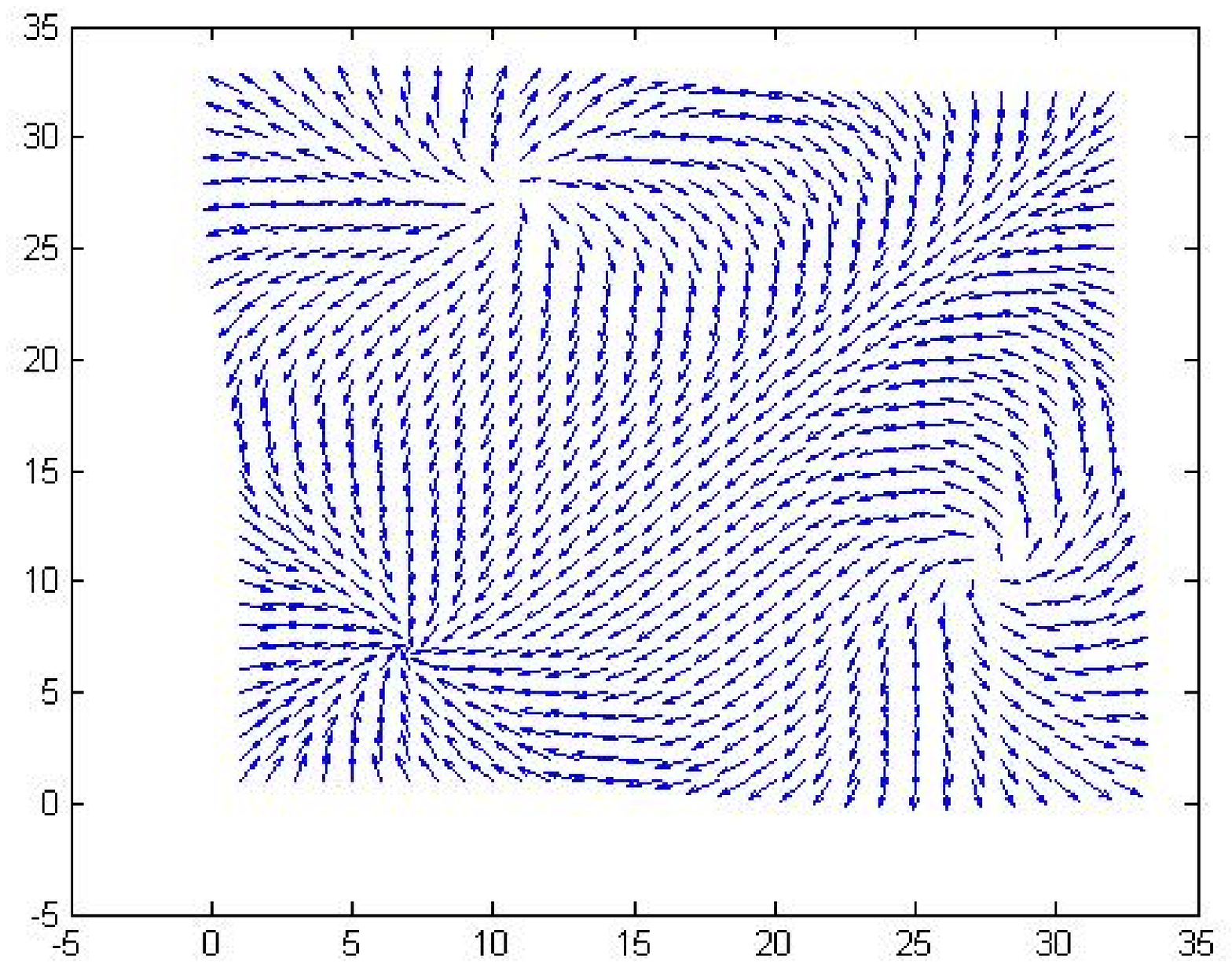}
\includegraphics[width=8cm,height=6cm]{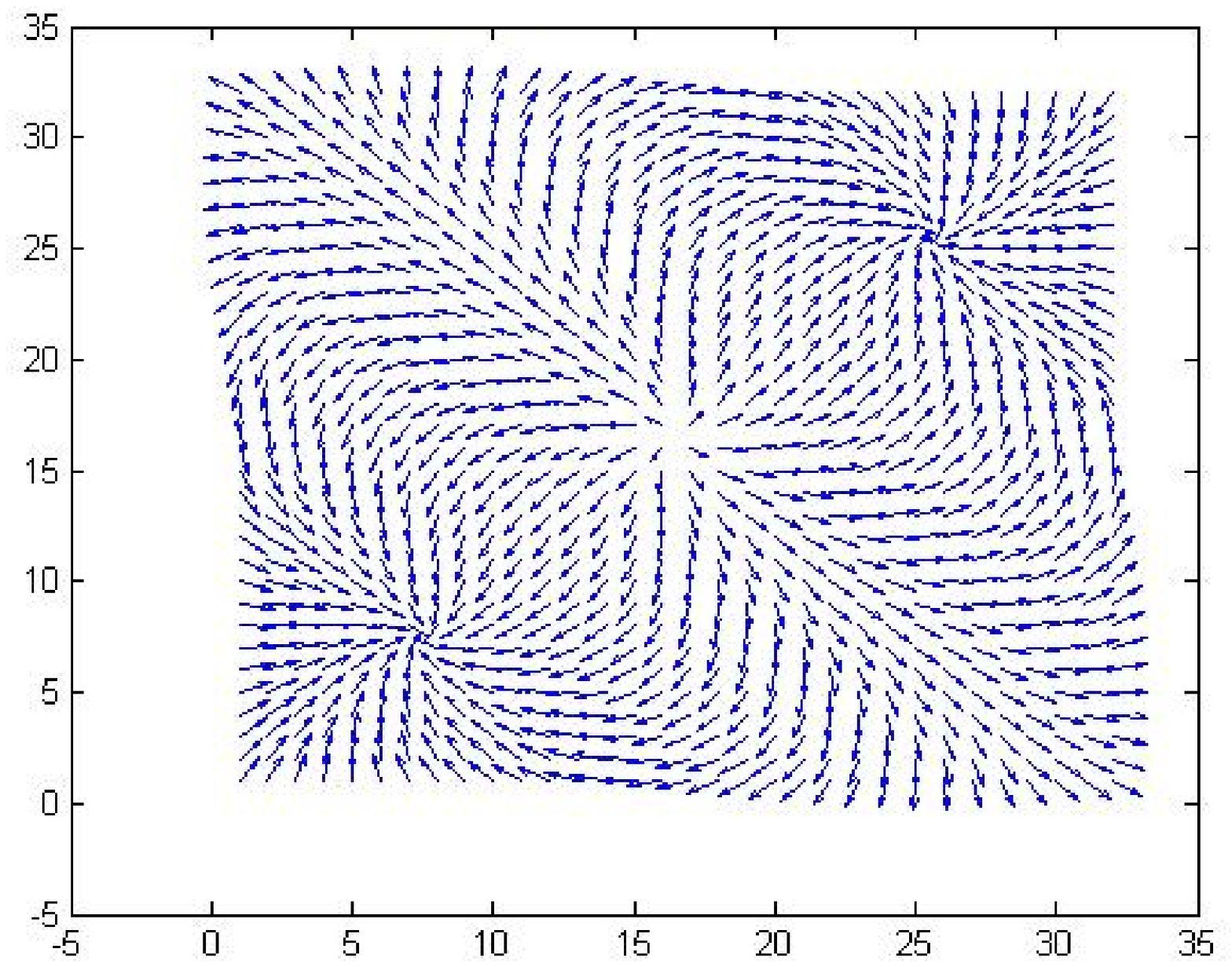}}
\leftline{{\sevenbf Fig 6.a}: $d_0=4, d_1=3, E=23$ \ \ \ \ \ \ \ \
\ \ \ \ \ \ \ \ \ \ \ \ \ \ \ \ \ \ \ \ \ \ {\sevenbf Fig 6.b}:
$d_0=-3, d_1=3, E=53$ }

\centerline{\includegraphics[width=8cm,height=6cm]{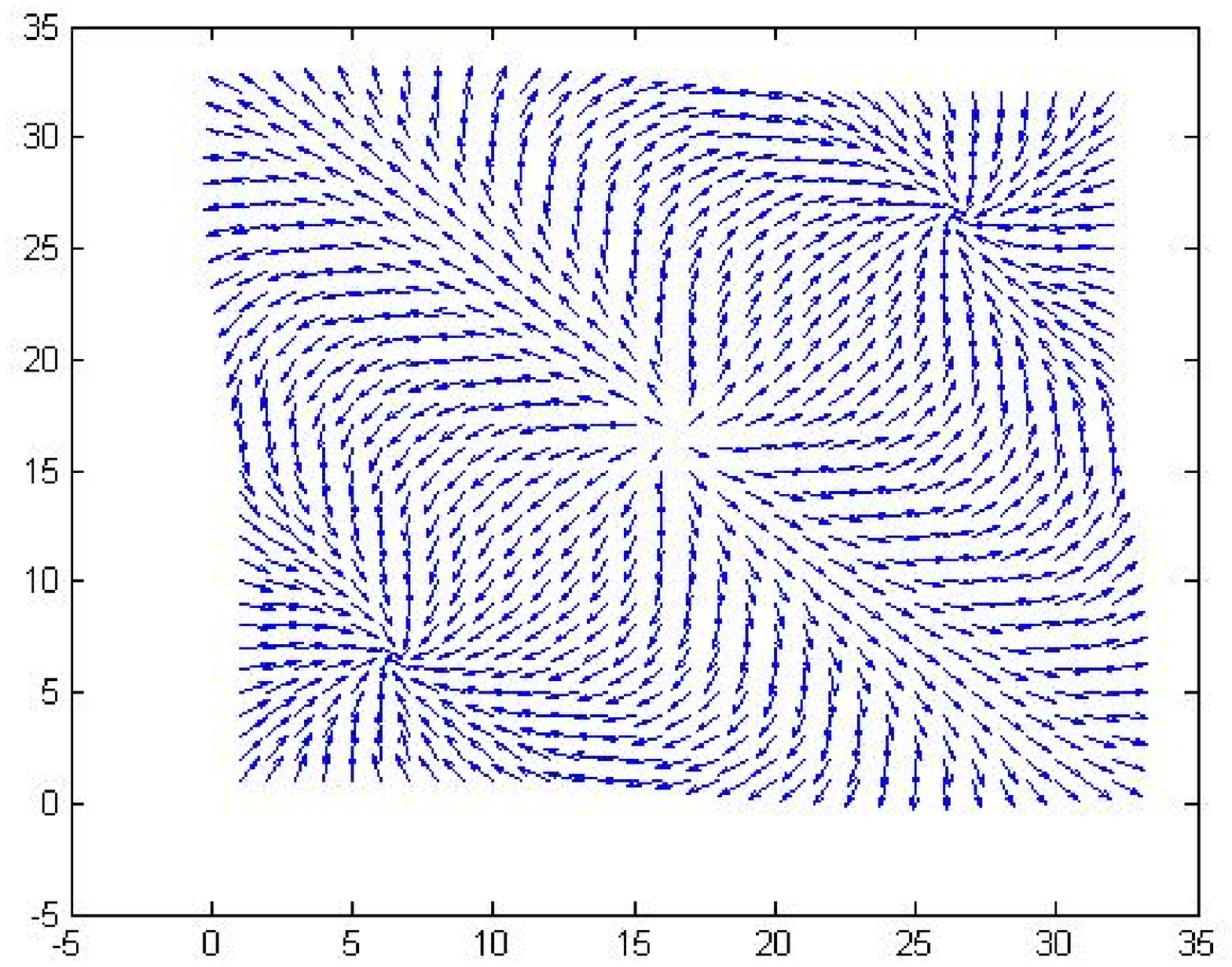}
\includegraphics[width=8cm,height=6cm]{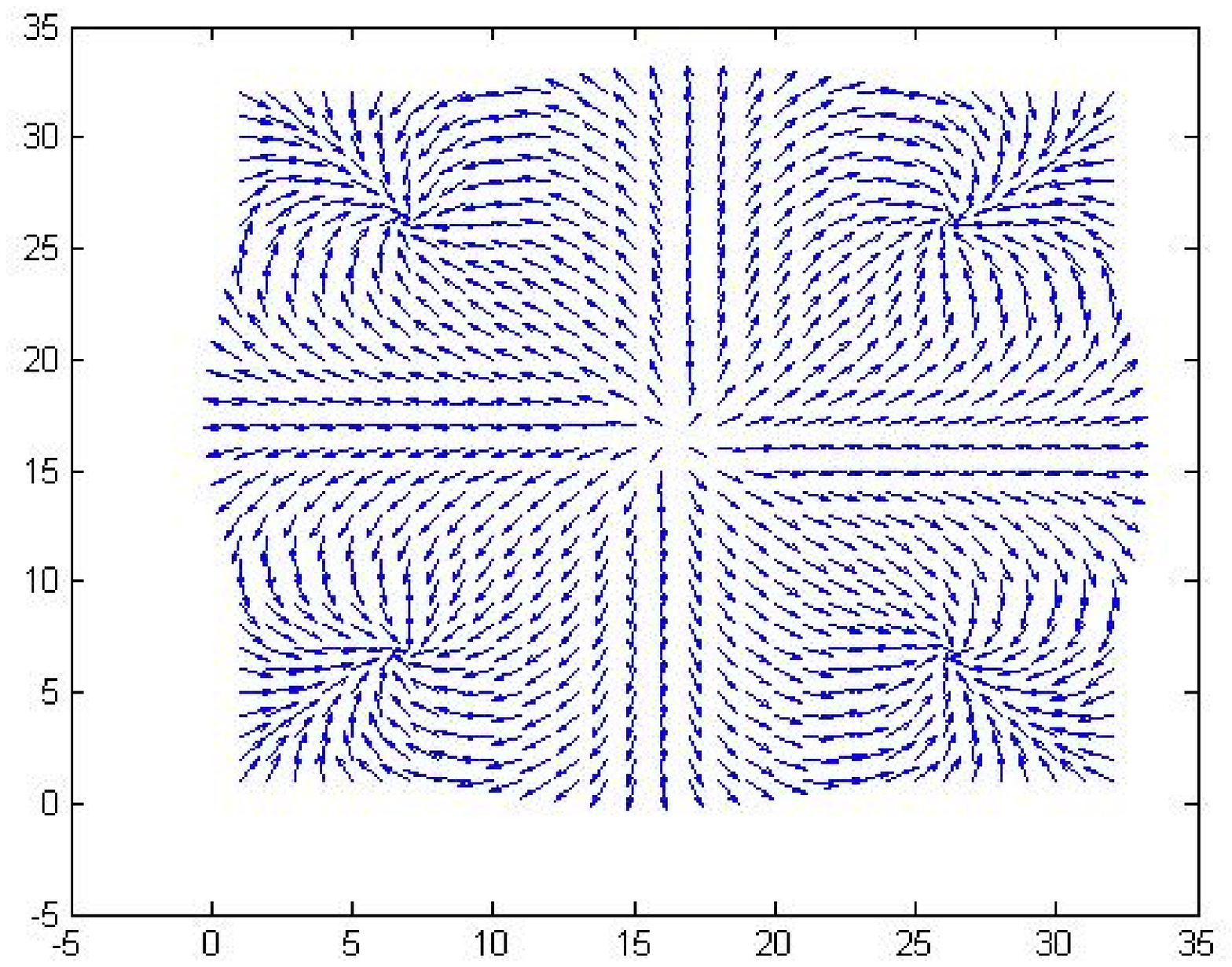}}
\leftline{\sevenrm {\sevenbf Fig 6.c}: $d_0=5, d_1=3, E=51$ \ \ \
\ \ \ \ \ \ \ \ \ \ \ \ \ \ \ \ \ \ \ \ \ \ \ \ \ \ \ \ \ \ \ \
{\sevenbf Fig 7.a}: $d_1=5, d=7$, zero initial condition}

\centerline{\includegraphics[width=8cm,height=6cm]{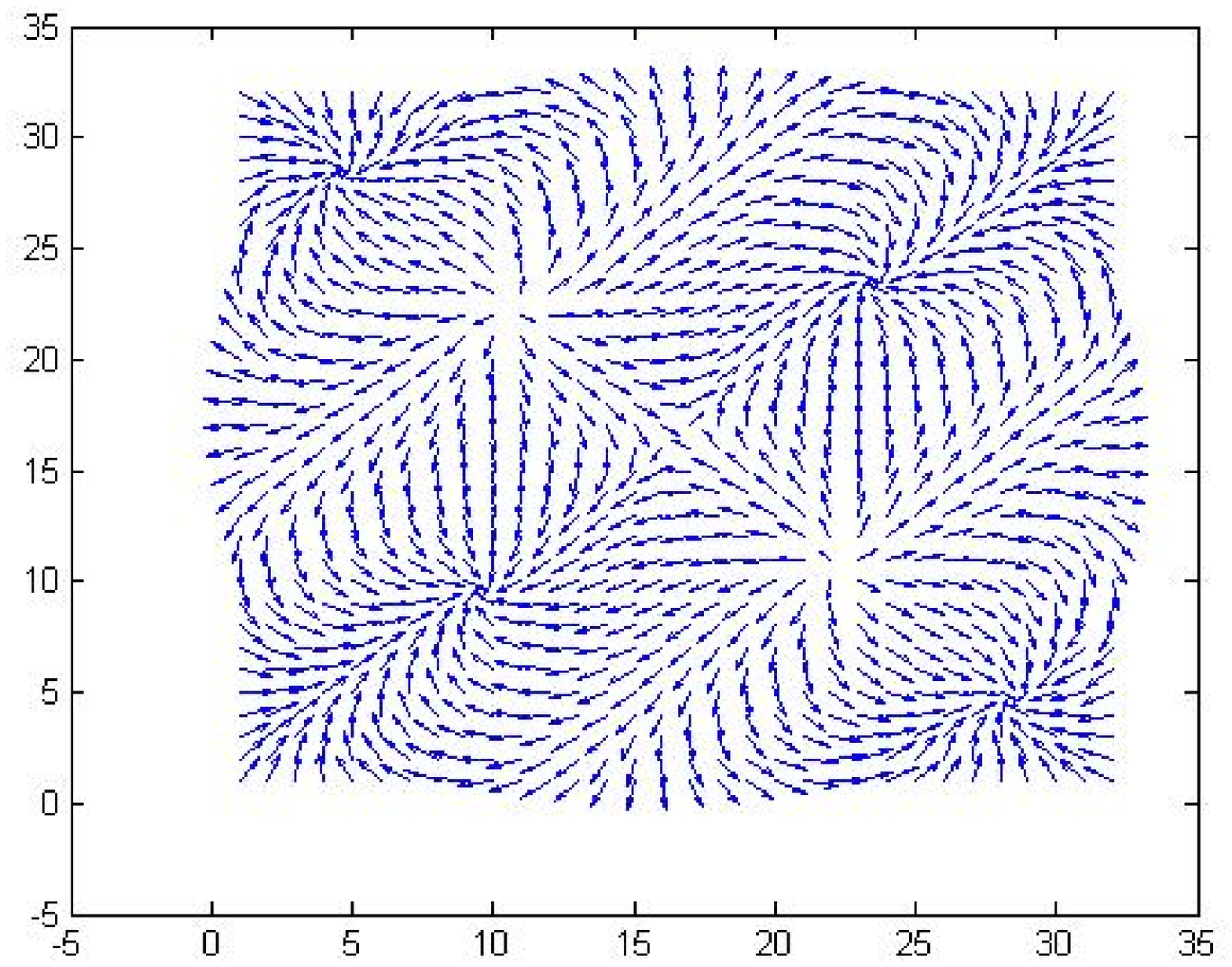}
\includegraphics[width=8cm,height=6cm]{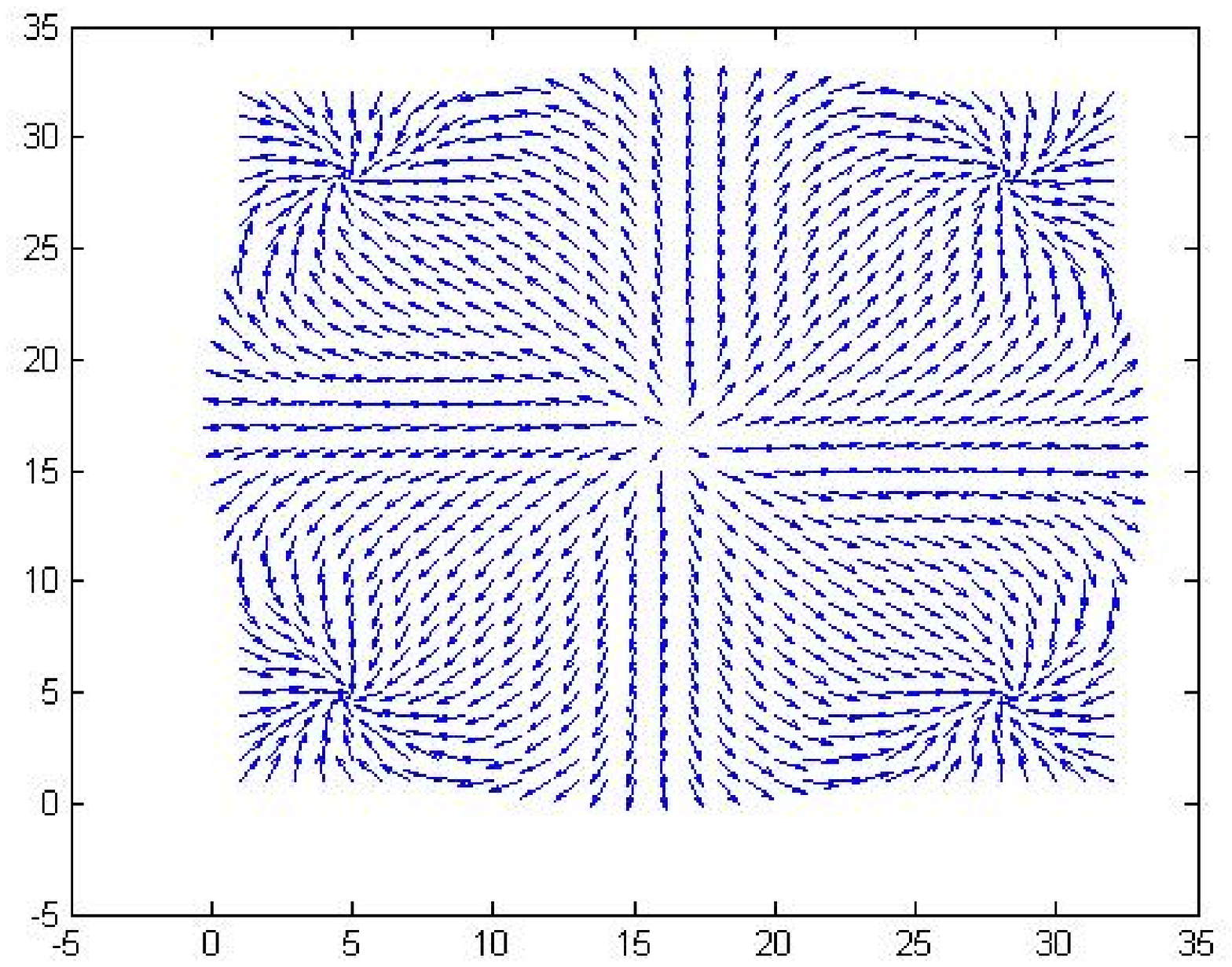}}
\leftline{\sevenrm {\sevenbf Fig 7.b}: $d_0=3, d_1=5$, without
annealing \ \ \ \ \ \ \ \ \ \ \ \ \ \ \ \ \ \ \ \ \ \ \ \ \
{\sevenbf Fig 7.c}: $d_0=3, d_1=5$, with annealing}

\medskip
In Fig. 7 we show how to pass from $d_0=3$ to $d_1=5$. The
configuration with zero initial condition and $d_1=5$ is given in
Fig. 5a, and energy is $E=113$. Taking instead the equilibrium
configuration for $d_0=3$ as an initial condition yields, without
simulated annealing, to Fig. 7b, with $d_1=6-1$, and $E=216$.
Using simulated annealing gives instead Fig. 7c, which looks like
Fig. 7a, and corresponding energy $E=115$. Actually, the 3
vortices on the anti-diagonal of the square in Fig. 7b collapse
into a single one at the center.

Note also that the degeneracy in case of $d_1=2$ (a vortex of
multiplicity 2 at the center for zero initial condition, $E=24$,~)
is lifted through annealing from $d_0=4$~: the 2 vortices move far
apart, and $E=-19$. Other applications of simulated annealing will
be given in the next subsection.

\medskip
\noindent{\bf d) More general configurations}.
\smallskip
We examine here the r\^ole of random fluctuations in the
distribution of spins on the boundary $\Lambda^{*c}$, so to
account for possible defects in the structure. With notations of
Sect.4a, we take $\sigma_j=\exp \bigl(2i\pi d
(j/N_\iota+\varepsilon_{\iota,j})\bigr)$, where
$\varepsilon_{\iota,j}$ are uniform i.i.d. random variables with
$\Sum_{j=1}^{N_\iota}\varepsilon_{\iota,j}=0$, and
$(\varepsilon_{\iota,j})_{\iota,1\leq j\leq N_\iota-1}$, and
variance small enough. The total degree is still equal to $d$, but
the variation of the direction of spins at the boundary is not
uniform. As expected, the picture does not depart drastically from
the previous cases. Vortices change their place according to the
initial value, and tend again to gather inside $\Lambda^*$, but
take always the value +1 (assuming $d>0$.) The sole effect of
randomness in the boundary condition is to change the place of the
vortices: namely they tend to get even closer to the boundary, so
to leave larger ordered regions in the middle.

In Fig 8.a,b, we have shown equilibrium configurations, obtained
for $d=7$, from the same initial and boundary conditions, but with
(resp. without) simulated annealing. Initial magnetization has
been chosen at random, but a priori larger than before, the sole
requirement being that $|m_0(x)|\leq m_\beta$. Random fluctuations
on the boundary have been prescribed as above.

\centerline{\includegraphics[width=8cm,height=6cm]{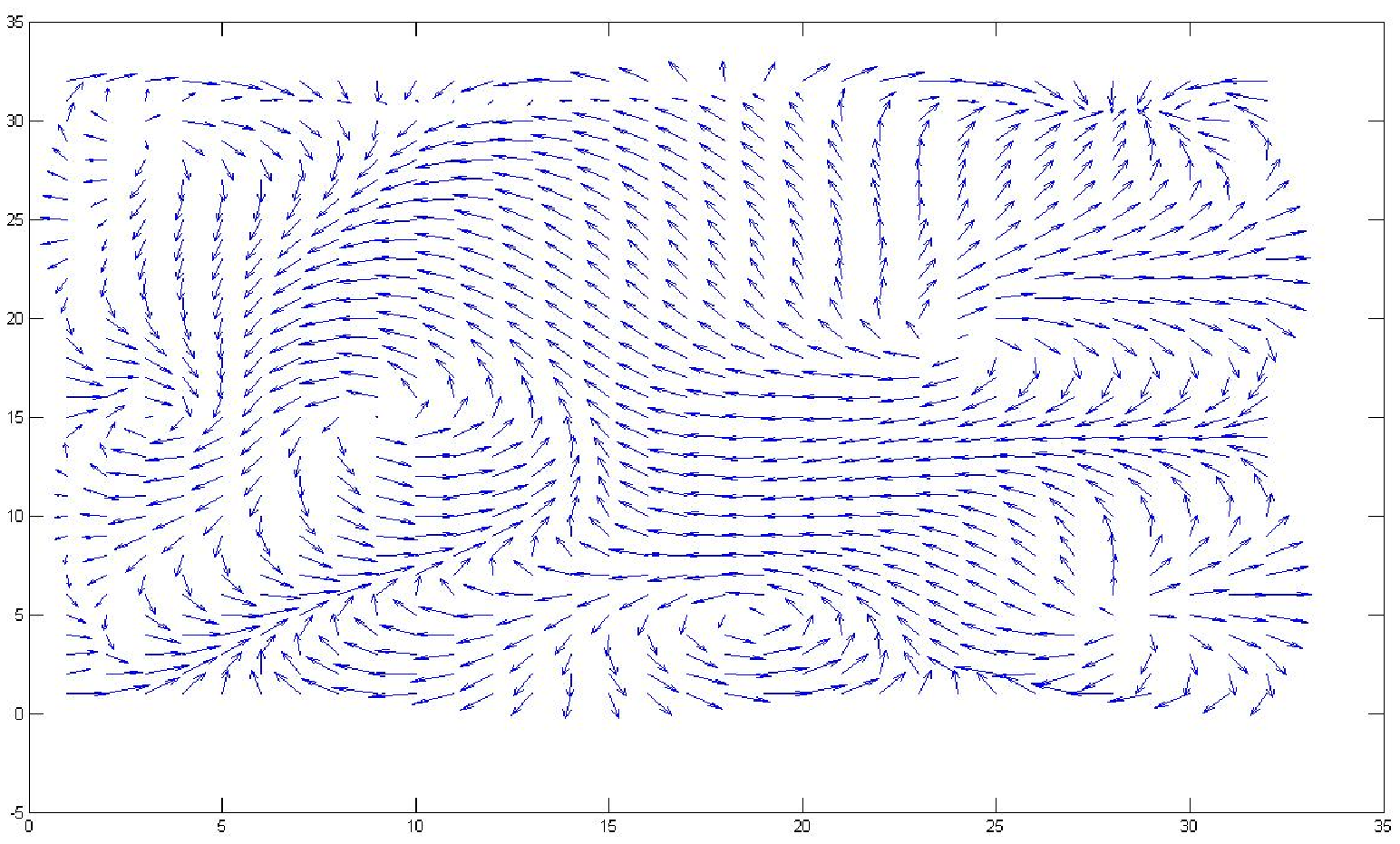}
\includegraphics[width=8cm,height=6cm]{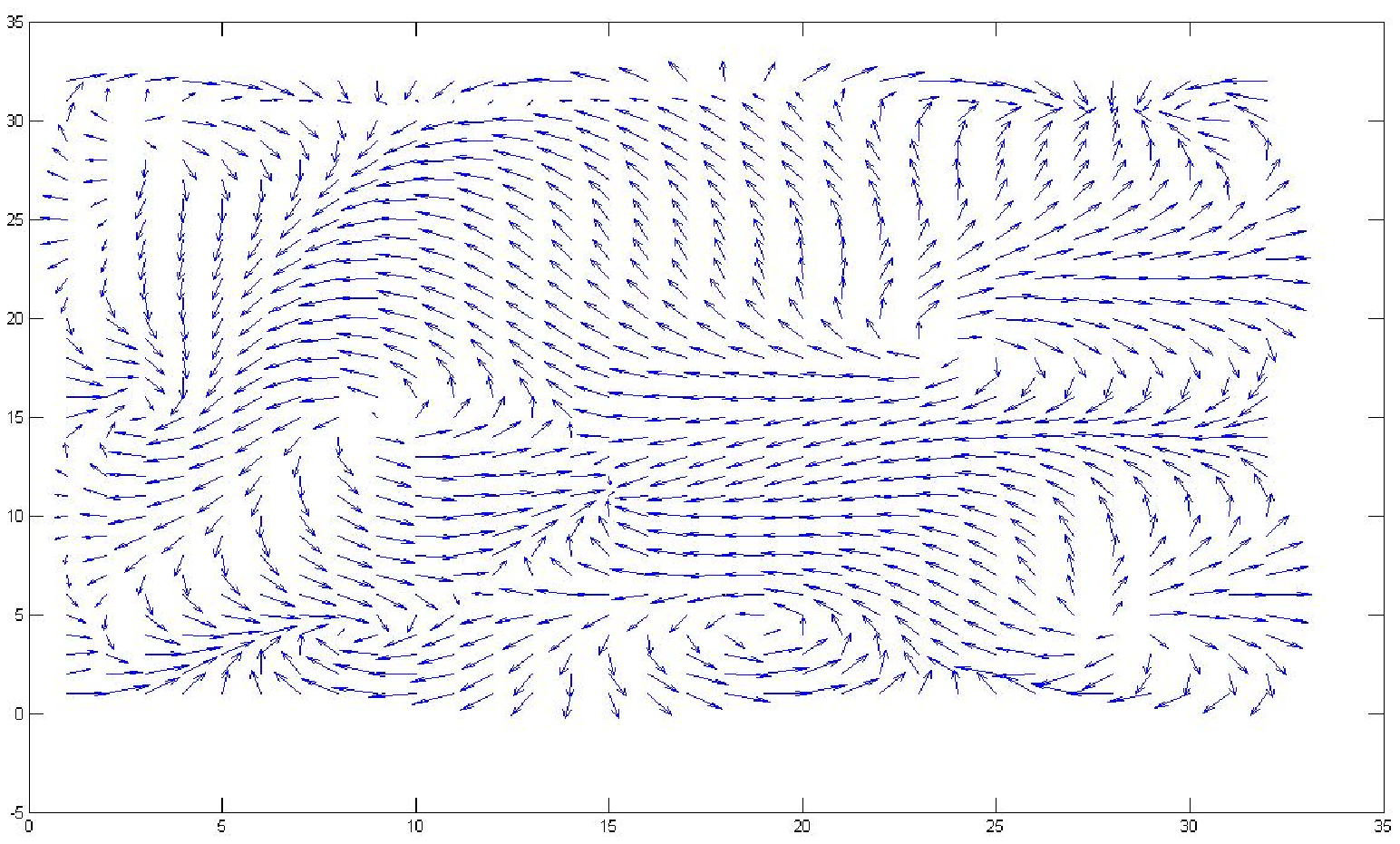}}
\leftline{\sevenrm {\sevenbf Fig 8.a}: $d=8-1, E=277$, without
annealing \ \ \ \ \ \ \ \ \ \ \ \ \ \ \ \ \ \ \ \ \ \ \ \
{\sevenbf Fig 8.b}: $d=7, E=240$, with annealing}

\medskip
\noindent{\bf e) The Kirchoff-Onsager correction}.
\smallskip
Another interesting result concerns the value of energy for the
minimizing configurations. In case of Ginzburg-Landau equation,
$-\Delta \psi+(|\psi|^2-1)\psi=0$, where $\psi$ is subject to a
boundary condition with vorticity, it is known that energy of the
minimizer vs. vorticity, has an asymptotic, as the $n$ vortices
$x_j$ become distant from each other, the leading order term is
given by a``proper energy'', proportional to $\Sum_{i=1}^nd_i^2$,
and the next correction is the inter-vortex energy given by
so-called Kirchhoff-Onsager hamiltonian, of the form
$$W_0=-\pi\Sum_{i\neq j}d_id_j\log|x_i-x_j|\leqno(4.5)$$
(see e.g. [BetBrHe] and [OvSi2] for precise statements.~) It can
be interpreted as the electrostatic energy for a system of charges
$d_j$ interacting through Coulomb forces. It turns out that,
despite forces in action have no electrostatic character,
Kirchhoff-Onsager correction holds with a good accuracy in our
case, even for long range interactions (i.e. for small $\gamma$,~)
but provided the inter-vortex distance is bounded below by the
range of the interaction. We have listed below some graphs of
$K={\cal F}(\cdot|m^c)-W_0$, obtained with uniform boundary
conditions, which show that $K$ roughly grows linearly with $d$
(cf Fig 9).

\centerline{\includegraphics[width=7cm,height=4cm]{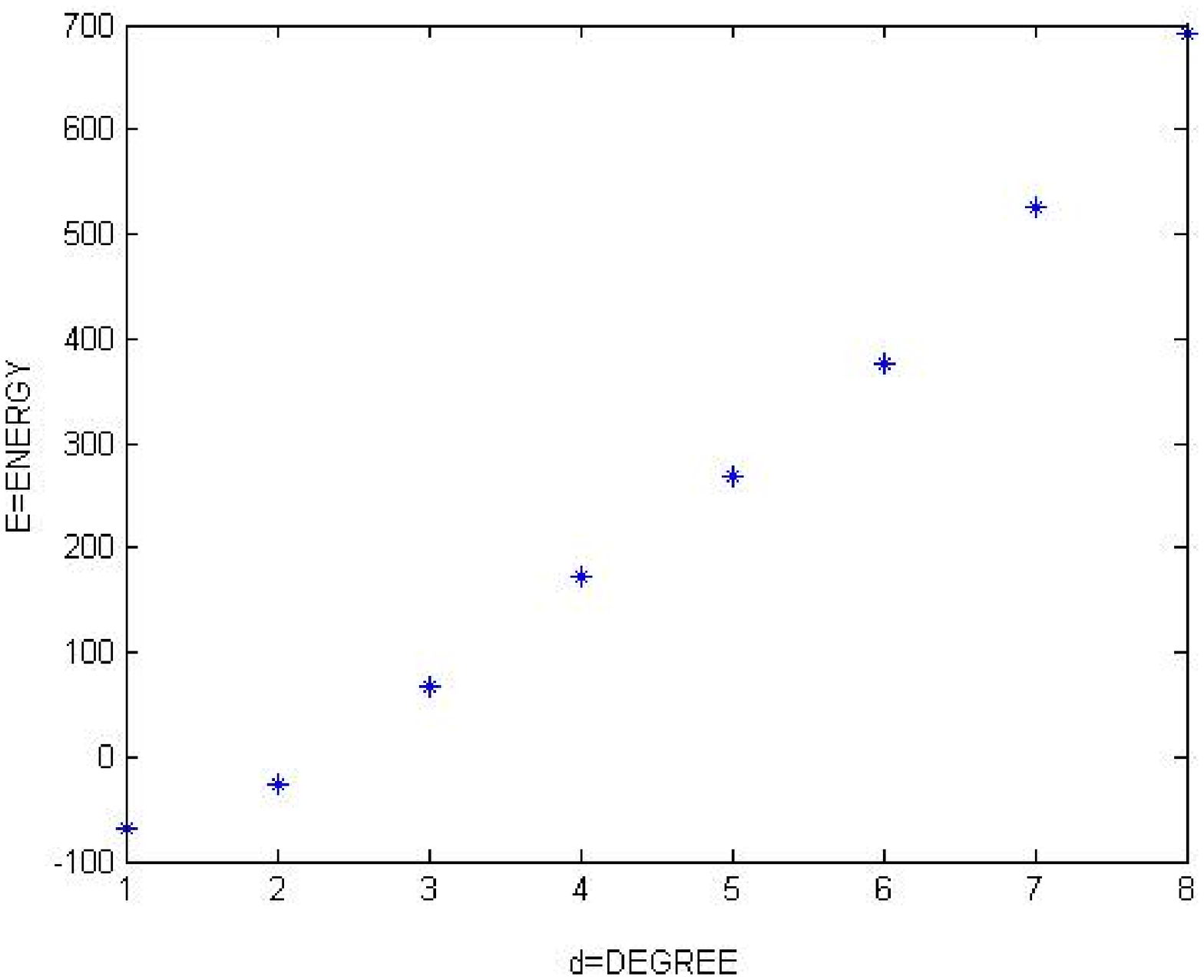}
\includegraphics[width=7cm,height=4cm]{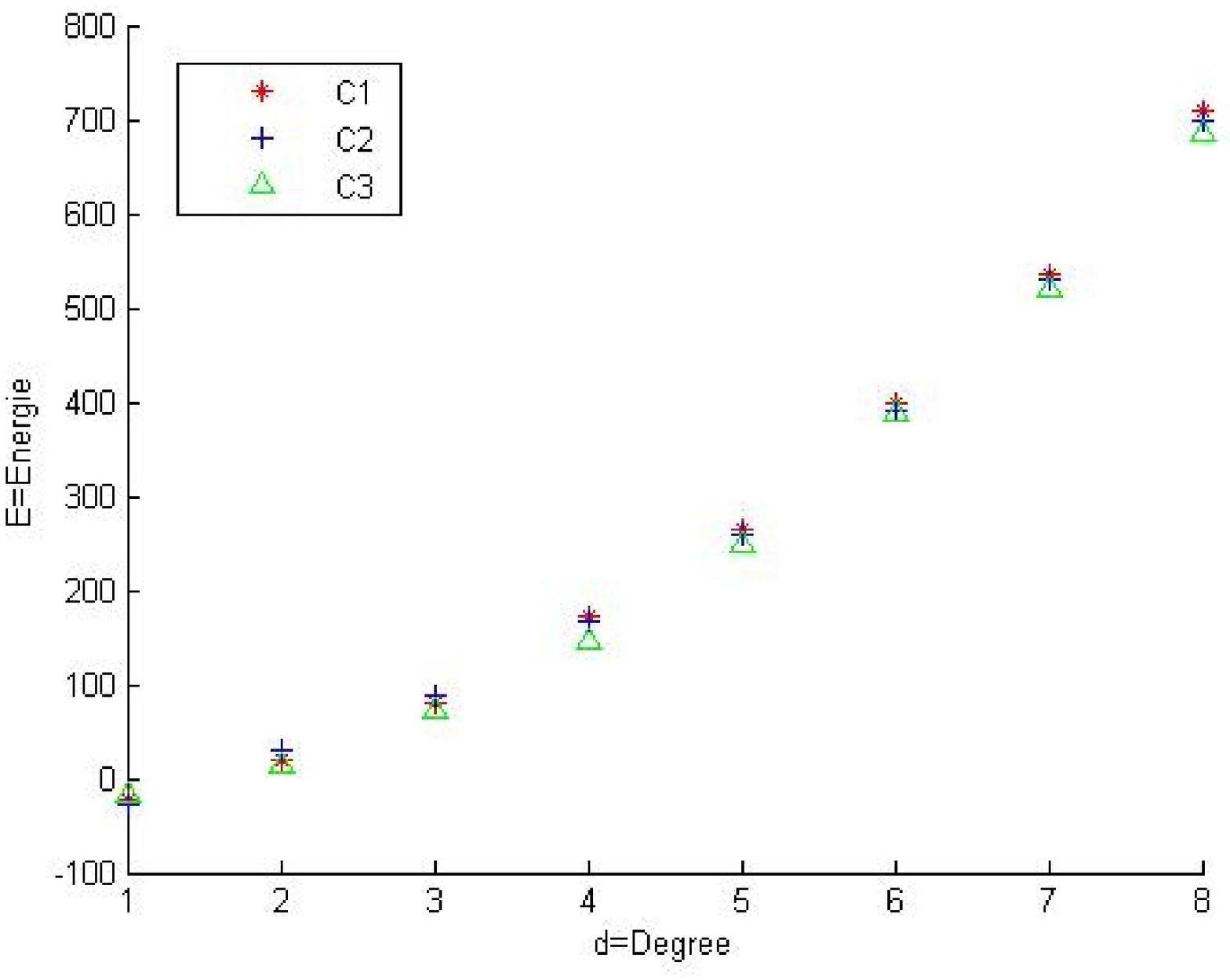}}
\centerline{\sevenrm \ \ {\sevenbf Fig 9.a}: $L^*=128$, zero
initial condition\ \ \ \ \ \ \ \ \ \ \ \ \ \ \ \ \ {\sevenbf Fig
9.b}: $L^*=128$, 3 random initial conditions}
\medskip

Fig 9.b shows that several  random trials for initial conditions
give approximately the same renormalized energy K.

\medskip
\noindent {\bf 5. The Heisenberg model}
\medskip
We consider here ``stationary spin waves'' for $q=3$, in a setting
similar to this of Belavin \& Polyakov [BePo], [Po,Chap.6].

Let us first recall the model. We look for minimizers of
$H(\sigma)=\int_{{\bf R}^2}|\nabla\sigma(x)|^2 dx$, among all
configurations $\sigma:{\bf R}^2\to{\bf S}^2$ subject to the
condition $\sigma(x)\to(0,0,1)$ as  $|x|\to\infty$.  This boundary
condition not only ensures a finite energy on the whole plane, but
also allows to extend $\sigma$ as a map on the one point
compactification ${\bf S}^2$ of ${\bf R}^2$, so  we may consider
its degree $D(\sigma)\in{\bf Z}$, or winding number, on the
sphere. Differentiable maps ${\bf S}^2\to{\bf S}^2$ are classified
by their degree, in the sense that $\sigma,\widetilde\sigma:{\bf
S}^2\to{\bf S}^2$ are homotopic iff they have the same degree. The
main result of Belavin and Polyakov asserts that there exist
solutions of that minimization problem, with given degree~; they
are called instantons, and expressed (in suitable coordinates
associated with the stereographic projection ${\bf C}\to{\bf
S}^2$) by arbitrary meromorphic functions of the form
$\prod_{j=1}^d{z-a_j\over z-b_j}$. Here $(a_j,b_j)\in{\bf C}^2$
play the role of vortices in the case $q=2$~; they have a natural
structure of dipoles, with poles placed at $a_j$ and $b_j$. So the
minimization problem (for a given homotopy class) has a continuous
degeneracy, parametrized by the family $(a_j,b_j)$ which we
interprete as moduli. The energy of all such instantons is a
constant proportional to $D$.

It is then natural to consider the contribution of all instantons
of same energy $D$. Somewhat heuristically, [Po] obtains, after
summing over $D\in{\bf N}$, a grand partition function of the form
$$
\Xi(\lambda)=\Sum_{D\geq 0}{\lambda^{2D}\over (D!)^2}\int\prod_j
da_jdb_j \exp\bigl[\Sum_{i<j}(\log |a_i-a_j|^2+\log
|b_i-b_j|^2)-\Sum_{i,j}\log|a_i-b_j|^2\bigr] \leqno(5.1)$$ and
each instanton behaves as if it consisted of a pair of opposite
Coulomb charges, placed at $a_j$ and $b_j$. Since the 2
dimensional Coulomb energy is given by $(1/4\pi)\log |a_j-b_j|^2$,
the exponent in (5.1) reminds us of the Kirchoff-Onsager
hamiltonian (4.5), and formally, $\Xi(\lambda)$ is the  grand
partition function of a plasma at inverse temperature
$\beta=4\pi$.

It is not known to which extend these instantons are stable
relatively to perturbations of $H(\sigma)$, e.g. due to the
influence of temperature.

We start with some considerations on the degree of a map on ${\bf
Z}^2$. Let $m:{\bf S}^2\to{\bf S}^2$ be a discrete map, defined
through the stereographic projection ${\bf Z}^2\to{\bf S}^2$, the
one point compactification of ${\bf Z}^2$ given  by $\overline{\bf
Z}^2\approx{\bf Z}^2\cup\{\omega\}$. The coordinates on the source
and target space are given by the polar and azimuthal angles
$(\theta,\varphi)$, and $(\widetilde\theta,\widetilde\varphi)$
respectively.

Consider the complex ${\cal C}= ({\bf Z}^2,L_{{\bf Z}^2},P_{{\bf
Z}^2})$ and its homology group. Here $L_{{\bf Z}^2}$ is the set of
bonds of unit length indexed by closest neighbors $x,x'\in{\bf
Z}^2$, and $P_{{\bf Z}^2}$ the set of chips  of unit area
(plaquettes) around $x\in{\bf Z}^2$. See e.g.[A] for concepts of
polyedral topology.

We define as usual the discrete jacobian $\Jac
m(x)={\partial(\widetilde\theta,\widetilde\varphi)\over
\partial(\theta,\varphi)}$ computed on the plaquette around $x$.
Let $y_0=m(x_0), x_0\in {\bf Z}^2$ be a regular value of $m$, i.e.
$\Jac m(x_0)\neq 0$.  The integer
$$D_{x_0}(m)=|\{x\in m^{-1}(\{y_0\}): \det \Jac m(x)>0\}|-|\{x\in
m^{-1}(\{y_0\}): \det \Jac m(x)<0\}|$$ is called local degree of
$m$ at $x_0$. In case where $D_{x_0}(m)$ takes the same value for
all $x_0\in {\bf Z}^2$, we call it the degree  of $m$  and denote
by $D(m)$. This is the general case, and $D(m)$ counts the number
of coverings of the sphere. Then $D(m)$ will be given by the
discrete analogue of the integral
$$D(m)={1\over 4\pi}\int_0^{2\pi}d\varphi\int_0^\pi d\theta\sin\theta
{\partial(\widetilde\theta,\widetilde\varphi),\over
\partial(\theta,\varphi)}$$
computed on the complex ${\cal C}$ defined above. When the values
of $m$  avoid a neighborhood of $\omega$, we put $D(m)=0$. If
$D_{x_0}(m)=d$ for all $x_0$ in a neighborhood of $\omega\in{\bf
Z}^2$, we call $d$ the degree of $m$ at infinity and denote
$d=D_\omega(m)$. See e.g. [BlGaRuSh] and references therein for a
more complete study of topological properties of discrete maps.

We conjecture that for Kac-Heisenberg model,  if $m$ is  a
minimizer for the free energy ${\cal F}(\cdot,|m^c)$, i.e. $m$
solves (3.2) or (3.3) with $D_\omega(m_0)=d$, after we take the
thermodynamical limit $\Lambda\to{\bf Z}^2$, then either $m$
vanishes at some point $x\in{\bf Z}^2$, or ${m\over |m|}:{\bf
Z}^2\to{\bf S}^2$ has degree $D$. In practice however, we have
only observed configurations with $0\leq D\leq d$. So $m$ shares
some features with Belavin \& Polyakov instantons, though with
less symmetries or degeneracies, and a possible ``degree loss''
from infinity, since we are not really working in the
thermodynamical limit.

It is straightforward to extend the constructions of Sect. 2 and 3
to the case $q=3$. Let us sketch the main steps. The moment
generating function is now $\phi(h)=\widehat\phi(|h|)={\sinh
|h|\over|h|}$, see [BuPi], and for the entropy function
$I(m)=\widehat I(|m|)$ defined in (2.2), we have $\widehat
I'=\bigl((\log\widehat\phi)'\bigr)^{-1}$, and
$(\log\widehat\phi)'(t)=L(t)={\cosh t\over\sinh t}-{1\over t}$
(the function $f$ before) is known as Langevin function. This is a
concave, increasing function on ${\bf R}^+$, $L(t)\sim t/3$ as
$t\to 0$, and $L(t)\to 1$ as $t\to\infty$. There is a phase
transition of mean field type i.e. a positive root for equation
$\beta m_\beta=\widehat I'(m_\beta)$, iff $\beta>\widehat
I''(0)=3$. We derive Euler-Lagrange equations for ${\cal
F}(m_\delta|m_\delta^c)$, as in Sect.3 (here we simply see $m$ as
a vector in the unit ball of ${\bf R}^3$, the complex
representation of $m$ was not essential,~) and find
$$-m+L(\beta|J_\delta *m|){J_\delta *m\over |J_\delta
*m|}=0\leqno(5.2)$$ For the corresponding gradient-flow dynamics
(3.4), there is again a free energy dissipation rate function,
which we compute exactly as in Proposition 3.1. Furthermore, we
have estimates on $m(x,t)$ as in Propositions 3.3 and 3.4~; more
precisely
\medskip
\noindent {\bf Proposition 5.1}: Assume $\beta> 3$, and let
$m(x,t)$ be the solution of (5.2) such that $m_0(x)=m(x,0)$
satisfies $|m_0(x)|\leq\lambda<1$, for some $\lambda\geq m_\beta$,
and all $x\in{\bf Z}^2$. Then $|m(x,t)|\leq\lambda$ for all
$x\in\Lambda^*$, and all $t>0$. Assume moreover the $z$-component
$m_0^z(x)$  of $m_0(x)$ satisfies $m_0^z(x))\geq\mu>0$, for all
$x\in{\bf Z}^2$, and some $\mu>0$ with
$(\mu^2+\lambda^2)^{1/2}<\beta L(\beta\lambda)$. Then
$m^z(x,t))\geq\mu$ for all $x\in\Lambda^*$, and $t>0$.

So choosing $\sigma^z(i)>0$ on $\Lambda^c$ (i.e. spins pointing to
the $z$ direction at the boundary) and also initial condition
$m_0^z(x)>0$ on $\Lambda^*$, Proposition 5.1 shows that
$m^z(t,x)>0$ stays bounded away from zero uniformly in time, so is
the case for the limiting orbit $m(x)$ on ${\bf Z}^2$, thus
$D(m)=0$. Our conjecture is again comforted by the following
numerical experiments, which also show that $m(x)$ depends in a
more essential way on the initial conditions than for the planar
rotator.
\medskip
\centerline{\includegraphics[width=8cm,height=6cm]{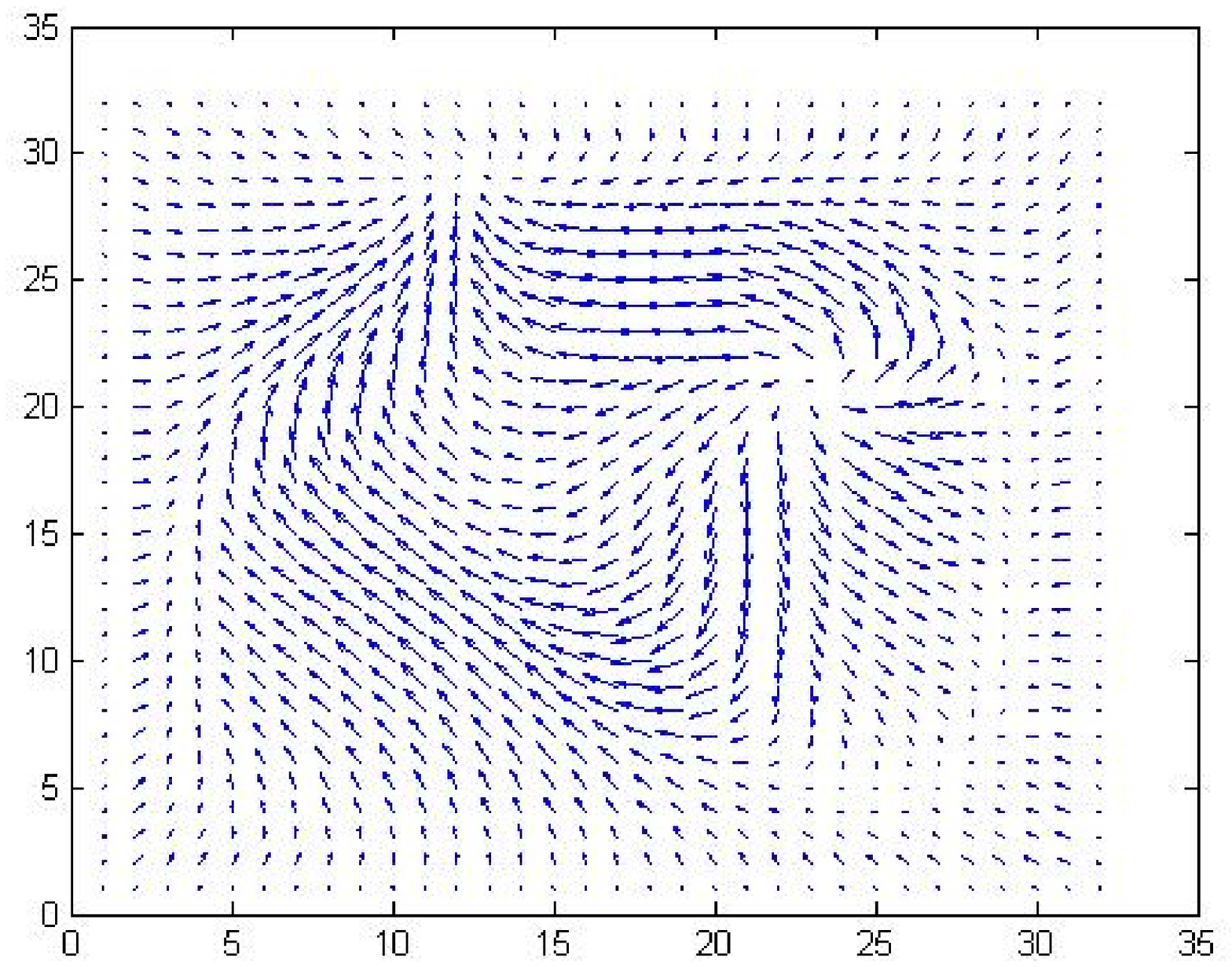}
\includegraphics[width=8cm,height=6cm]{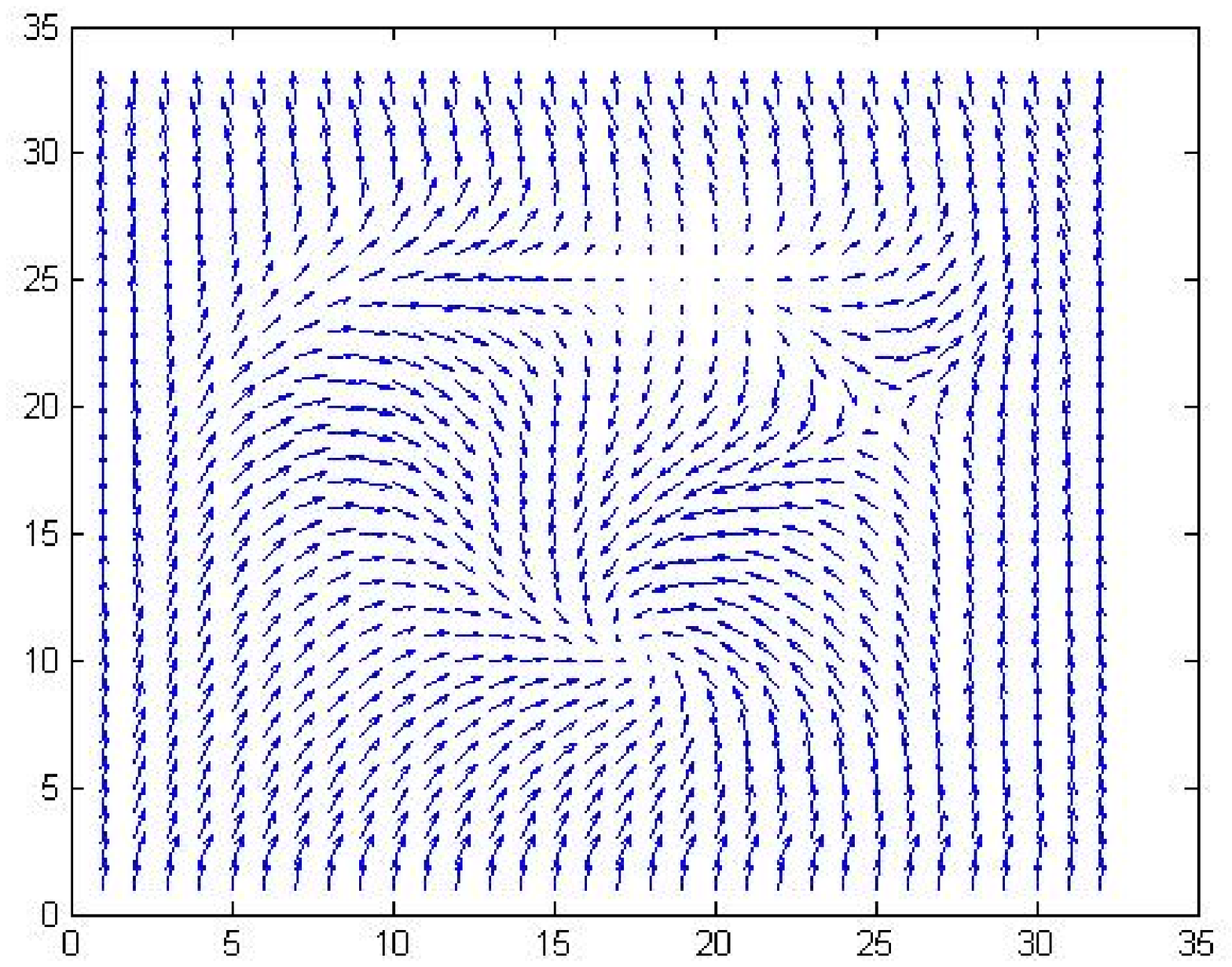}}
\centerline{\sevenrm {\sevenbf Fig 10.a}: $L^*=128, d=1$, XY plane
\ \ \ \ \ \ \ \ \ \ \ \ \ \ \ \ \ \ \ \ \ \ \ \ \ \ \ \ \ \
{\sevenbf Fig 10.b}: $L^*=128, d=1$, YZ plane}

\vskip 1truecm
We start with prescribing the spins variables on
$\Lambda^c$ as in (4.3), taking a family of loops
$\Gamma_i\subset\Lambda^c$, $i=0,1,2\cdots$ along which
$\sigma_j=(\cos\Phi_{ij}\sin\theta_i, \sin\Phi_{ij}\sin\theta_i,
\cos\theta_i)$, $\Phi_{ij}=2\pi dj/N_i+\phi_0$, and
$0\leq\theta_i\leq\theta_0$, decreasing with $i$, $\theta_{i_0}=0$
on the last loop $\Gamma_{i_0}$ interacting with $\Lambda$, and
$\theta_0$ small enough to fit with Belavin-Polyakov conditions.
So fixing the precession number $d=D_\omega(m)$, we get a
``stationary spin wave pattern'' on the boundary. Inside
$\Lambda^*$ we choose random initial values, $|m_0(x)|\leq
m_\beta$.

\vskip 1truecm
\centerline{\includegraphics[width=8cm,height=6cm]{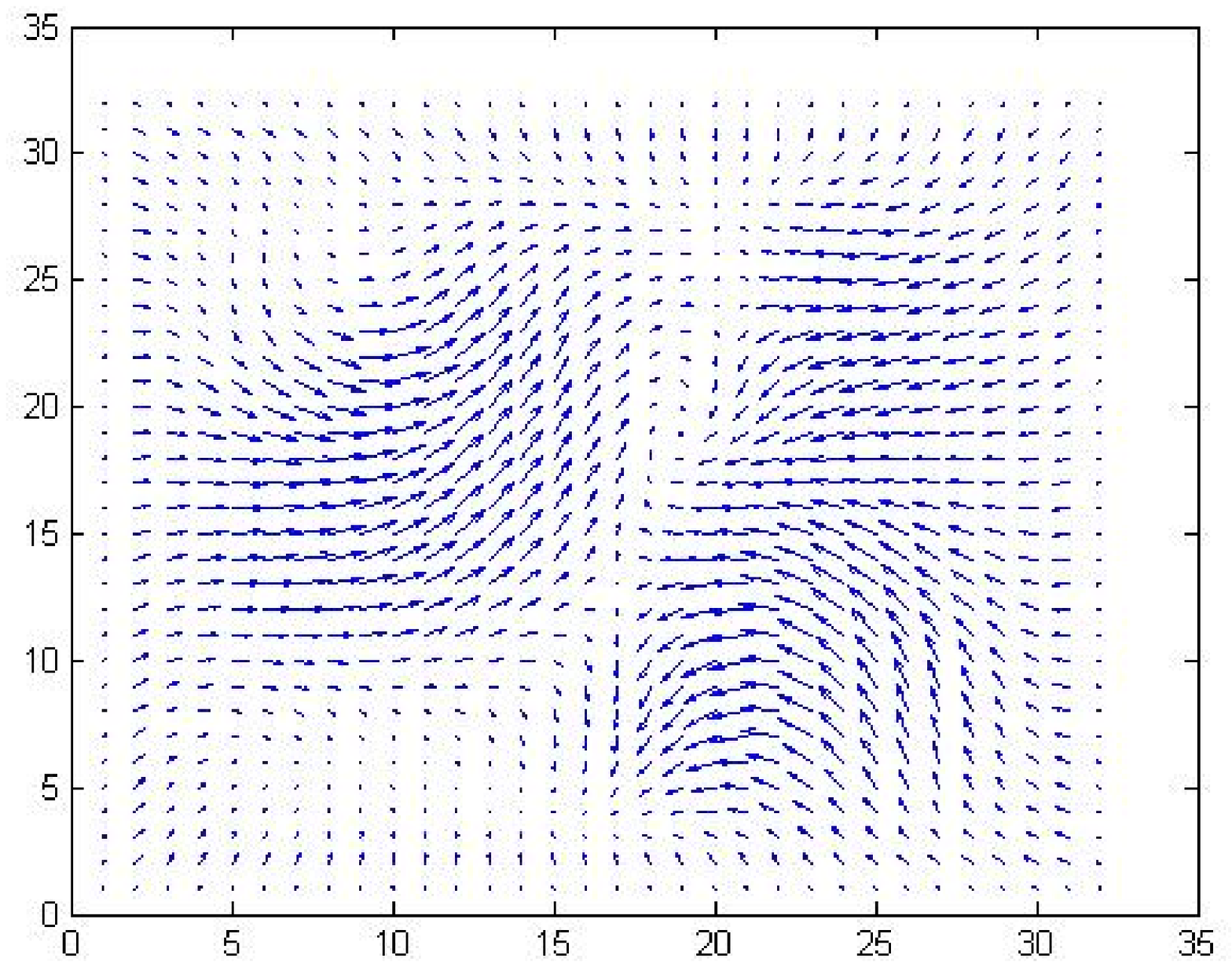}
\includegraphics[width=8cm,height=6cm]{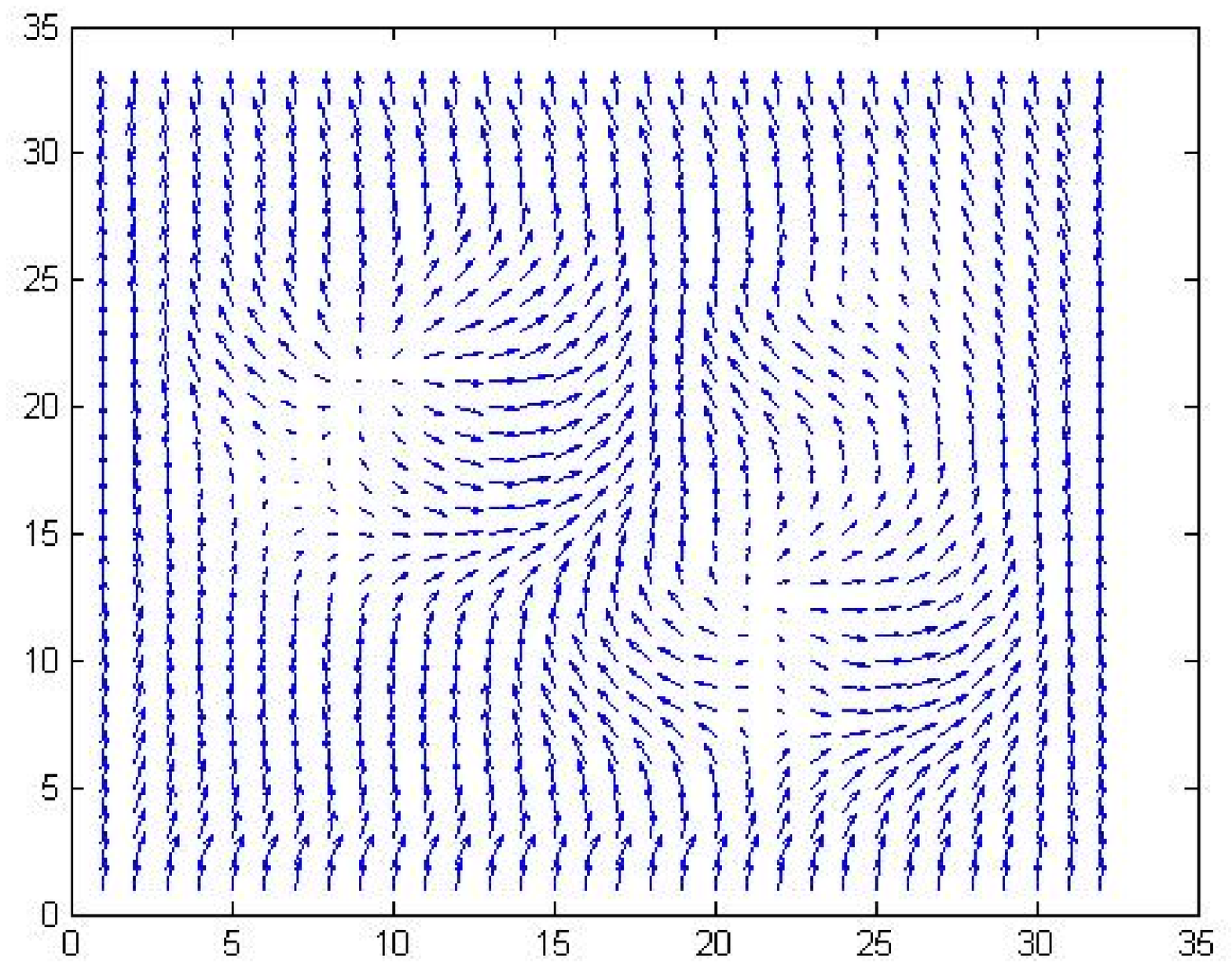}}
\centerline{\sevenrm {\sevenbf Fig 11.a}: $L^*=128, d=1$, XY
plane\ \ \ \ \ \ \ \ \ \ \ \ \ \ \ \ \ \ \ \ \ \ \ \ \ \ \ \ \
{\sevenbf Fig 11.b}: $L^*=128, d=1$, YZ plane}

We represent here a few sample of $(x,y)$ and $(y,z)$ projections
of the field $m$, which yield the following observations. In
general, the solution is very sensitive to the choice of initial
conditions, and many patterns show up, which reflects the moduli
in Belavin-Polyakov model. For relatively small $\beta$ (e.g.
$\beta=5$ with $L^*=128$) spin waves fluctuate, and $m^z$ can take
negative values, but the  domain where spins point downwards is
not sufficiently large to start revolving around the sphere.  So
the degree is $D=0$. We still get 2 dimensional ``vortices'' in
the $(x,y)$ plane, there are typically 1, 2 or 3 such
``vortices''when $d=1$, and up to 4 when $d=2$.  Exceptionally, we
can also get a 3 dimensional vortex, i.e. $x_0$ such that $m(x_0)$
becomes quite small. Such a  $m$ is no longer homotopic to a
function on the sphere.

Increasing $\beta$ generally prevents getting too small values for
$m$, and allows larger negative $m^z$. For $\beta=10$ and
$L^*=128$, there are random trials where the winding number $D$ is
non zero. Thus Fig.10 is obtained for $d=1$ and suggests also
$D=1$. In Fig.11 we still have $d=1$, but $D=0$, although 2 large
symmetric regions contain negative values of $m^z$. Fig.12 gives
an example where $D=1$ for $d=2$. Nevertheless, we have not
observed winding numbers $D=2$ for $d=2$.

\centerline{\includegraphics[width=8cm,height=6cm]{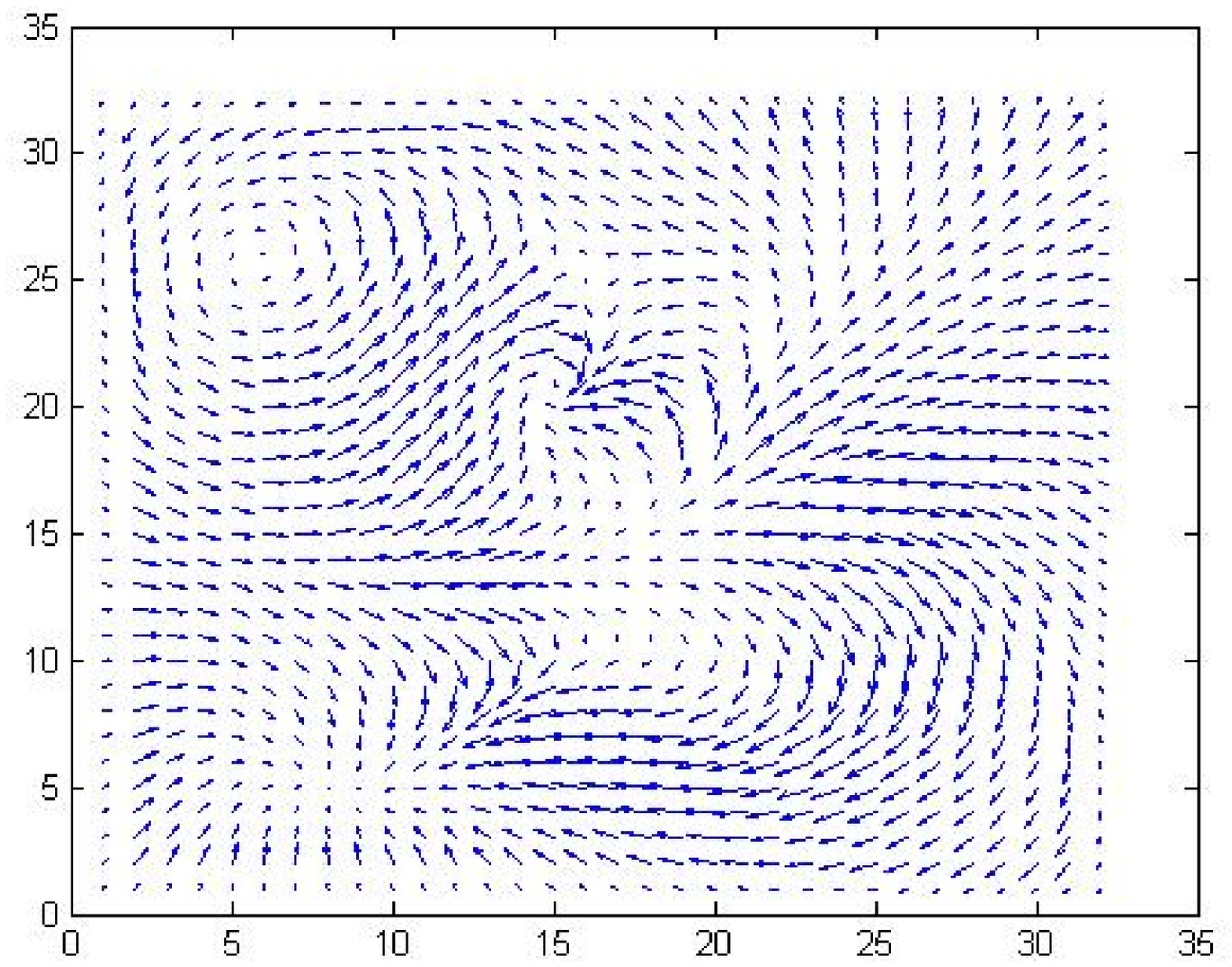}
\includegraphics[width=8cm,height=6cm]{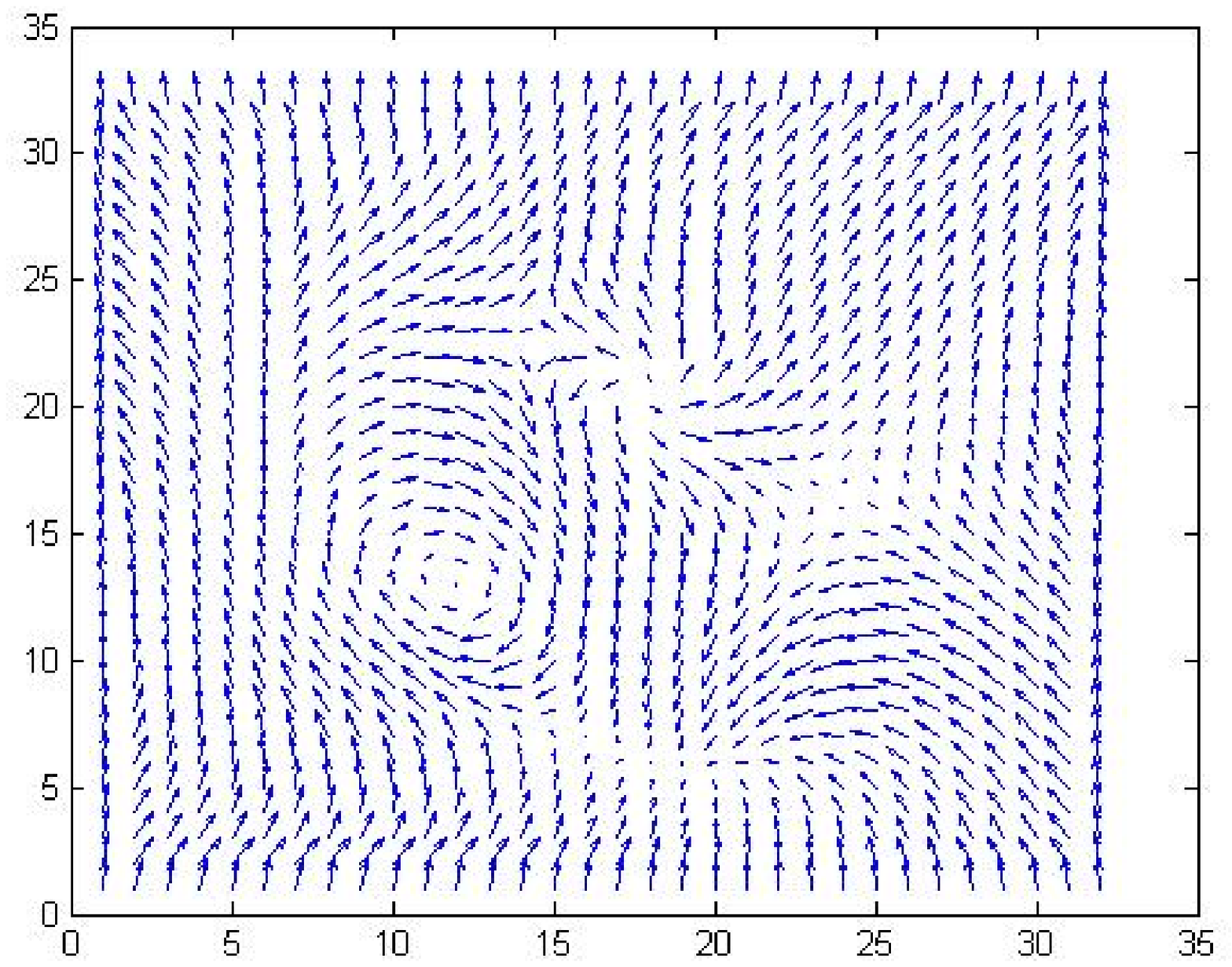}}
\centerline{\sevenrm {\sevenbf Fig 12.a}: $L^*=128, d=2$, XY
plane\ \ \ \ \ \ \ \ \ \ \ \ \ \ \ \ \ \ \ \ \ \ \ \ \ \ {\sevenbf
Fig 12.b}: $L^*=128, d=2$, YZ plane}

\vskip 1truecm

Further increasing $\beta$ for a given $L^*$ doesn't reveal
anything new~; namely, if small temperature seems to favors long
range order and existence of non trivial instantons, it also
creates stiffness and a need for space. In any case, one should
keep in mind that Belavin-Polyakov instantons can be reproduced
only as $\beta\to \infty$, and in the thermodynamic limit
$|\Lambda|\to\infty$. Of course, everything can be again improved
through simulated annealing.

\medskip
\centerline{\bf References}
\medskip
\noindent [A] P.Alexandroff. Elementary Concepts of Topology.
Dover Publ. N.Y., 1961.

\noindent [AlBeCaPr] G.Alberti, G.Belletini, M.Cassandro,
E.Presutti. Surface tension in Ising systems with Kac potentials.
J. Stat. Phys. Vol.82, (3 and 4) 1996, p.743-795.

\noindent [BetBrHel] F.Bethuel, H.Brezis, F.Helein.
Ginzburg-Landau vortices, Birkh\"auser, Basel 1994.

\noindent [BePo] A.A.Belavin, A.M.Polyakov. Metastable states of
2-d isotropic ferromagnets JETP Lett., Vol 22, No.10,1975,
p.245-247.

\noindent [BlGaRuSh] Ph.Blanchard, D.Gandolfo, J.Ruiz, and
S.Shlosman. On the Euler-Poincar\'e characteristic of random
cluster model. Mark. Proc. Rel. Fields, No.9, 2003, p.523.

\noindent [BleLe] P.Bleher, J.Lebowitz. Energy-level statistics of
model quantum systems: universality and scaling in a lattice-point
problem. J.Stat.Phys. 74, 1994, p.167-217.

\noindent [BuPi] P.Butt\`a, P.Picco. Large-deviation principle for
one-dimensional vector spin models with Kac potentials.
J.Stat.Phys. 92, 1998, p.101-150.

\noindent [DeM] A.DeMasi. Spins systems with long range
interactions. Progress in Probability, Birkh\"auser, Vol 54, 2003,
p.25-81.

\noindent [DeMOrPrTr] A.DeMasi, E.Orlandi, E.Presutti, L.Triolo.
Uniqueness and global stability of the instanton in non-local
evolution equations. Rendiconti di Mat., Serie VII, 14, 1994,
p.693-723.

\noindent [El-BoRo] H.El-Bouanani, M.Rouleux. Thermodynamical
equilibrium of vortices for the bidimensional continuous Kac
rotator. In preparation.

\noindent [H\"o] L.H\"ormander. The Analysis of Partial
Differential Operators I. Springer, 1983.

\noindent [KiGeVec] S.Kirkpatrick, C.D.Gelatt, M.P.Vecchi.
Optimization by simulated annealing. Science, Nr. 4598, 1983.

\noindent [LeVeRu] X.Leoncini, A.Verga, S.Ruffo. Hamiltonian
dynamics and the phase transition of the XY model. Phys. Rev. E
57, 1998, p.6377-6389.

\noindent [MiZh] R.Minlos, E.Zhizhina. Asymptotics of the decay of
correlations for the Gibbs spin fields.  Theoret.Math.Phys. 77(1),
1988, p.1003-1009.

\noindent [OvSi] Y.Ovchinnikov, I.M.Sigal. Ginzburg-Landau
Equation I. Static vortices. CRM Proceedings. Vol 12, 1997,
p.199-220.

\noindent [OvSi2] Y.Ovchinnikov, I.M.Sigal. The energy of
Ginzburg-Landau vortex, European J. of Applied Mathematics 13,
2002, p.153-178.

\noindent [Po] A.M.Polyakov. Gauge fields and strings. Harwood
Academic. Chur. 1987.

\noindent [Pr] E.Presutti. From statistical mechanics towards
continuum mechanics. Preprint M.Planck Institute, Leipzig, 1999

\noindent [Ru] D.Ruelle. Statistical Mechanics. World Scientific,
1999.

\noindent [S] Y.Sinai. Theory of phase transitions: rigorous
results. Pergamon Press, 1982.

\noindent [Z] J.Zinn-Justin. Quantum Fields Theory and Critical
Phenomena. Clarendon Press, Oxford, 1989.

\bye

\bye